\documentclass[12pt,reqno]{amsart}
\theoremstyle{plain}
\newtheorem{thm}{Theorem}
\theoremstyle{plain}
\newtheorem{cor}{Corollary}
\theoremstyle{plain}
\newtheorem{lem}{Lemma}
\theoremstyle{plain}
\newtheorem{prop}{Proposition}
\theoremstyle{definition}
\newtheorem*{defn*}{Definition}
\theoremstyle{remark}
\newtheorem*{rem*}{Remark}
\theoremstyle{remark}
\newtheorem*{rems*}{Remarks}

\newcounter{cps}
\newcommand{\thmcount}{\setcounter{thm}{\value{cps}}
   \addtocounter{cps}{1}}
\newcommand{\propcount}{\setcounter{prop}{\value{cps}}
   \addtocounter{cps}{1}}
\newcommand{\lemcount}{\setcounter{lem}{\value{cps}}
   \addtocounter{cps}{1}}
\newcommand{\corcount}{\setcounter{cor}{\value{cps}}
   \addtocounter{cps}{1}}

\DeclareMathOperator{\diag}{diag}
\DeclareMathOperator{\dist}{dist}
\DeclareMathOperator{\Spec}{Spec}
\DeclareMathOperator{\Tr}{Tr}
\DeclareMathOperator{\Map}{Map}
\DeclareMathOperator{\Ran}{Ran}
\DeclareMathOperator{\sgn}{sgn}
\DeclareMathOperator{\Dom}{Dom}
\DeclareMathOperator{\ad}{ad}
\DeclareMathOperator{\Ker}{Ker}
\DeclareMathOperator{\Li}{Li}
\DeclareMathOperator{\const}{const}

\newcommand{\C}{\mathbb{C}}
\newcommand{\N}{\mathbb{N}}
\newcommand{\R}{\mathbb{R}}
\newcommand{\Z}{\mathbb{Z}}

\newcommand{\bA}{\mathbf{A}}
\newcommand{\bB}{\mathbf{B}}
\newcommand{\bC}{\mathbf{C}}
\newcommand{\bD}{\mathbf{D}}
\newcommand{\bK}{\mathbf{K}}
\newcommand{\bU}{\mathbf{U}}
\newcommand{\bV}{\mathbf{V}}
\newcommand{\bW}{\mathbf{W}}
\newcommand{\bX}{\mathbf{X}}
\newcommand{\bk}{\mathbf{k}}
\newcommand{\br}{\mathbf{r}}

\newcommand{\BB}{\mathcal{B}}
\newcommand{\DD}{\mathcal{D}}
\newcommand{\HH}{\mathcal{H}}
\newcommand{\UU}{\mathcal{U}}
\newcommand{\VV}{\mathcal{V}}

\newcommand{\XX}{\mathcal{X}}
\newcommand{\KK}{\mathcal{K}}

\newcommand{\gA}{\mathfrak{A}}
\newcommand{\gX}{\mathfrak{X}}
\newcommand{\gY}{\mathfrak{Y}}

\renewcommand{\i}{\imath}

\newcommand{\Omegabad}{\Omega^{\mathrm{bad}}}

\begin{document}
\title{Weakly regular Floquet Hamiltonians\\ with pure point spectrum}
\author{P. Duclos$\ ^{1,2}$, O. Lev$\ ^{3}$, P.
\v{S}\v{t}ov\'i\v{c}ek$\ ^{3}$,
M. Vittot$\ ^{1} $}
\maketitle
\thispagestyle{empty}
{\raggedright
\( ^{1}\ \)Centre de Physique Th\'eorique, CNRS, Luminy, Case 907,
13288 Marseille Cedex 9,\\ France

\vspace{0.1cm}
\( ^{2}\ \)PHYMAT, Universit\'e de Toulon et du Var, BP 132, F-83957
La Garde Cedex, France

\vspace{0.1cm}
\( ^{3}\ \)Department of Mathematics, Faculty of Nuclear Science,
Czech Technical University, Trojanova 13, 120 00 Prague, Czech
Republic
}

\begin{abstract}
We study the Floquet Hamiltonian \( -i\partial _{t}+H+V(\omega t) \),
acting in \linebreak \( L^{2}([\, 0,T\, ],\HH ,dt) \), as depending
on the parameter \( \omega =2\pi /T \). We assume that the spectrum
of \( H \) in \( \HH  \) is discrete, \( \Spec
(H)=\{h_{m}\}_{m=1}^{\infty } \),
but possibly degenerate, and that \( t\mapsto V(t)\in \BB (\HH ) \)
is a \( 2\pi  \)-periodic function with values in the space of
Hermitian
operators on \( \HH  \). Let \( J>0 \) and set \( \Omega _{0}=[\,
\frac{8}{9}J,\frac{9}{8}J\, ] \).
Suppose that for some \( \sigma >0 \) it holds true that \( \sum
_{h_{m}>h_{n}}\mu _{mn}(h_{m}-h_{n})^{-\sigma }<\infty  \)
where \( \mu _{mn}=(\min \{M_{m},M_{n}\})^{1/2}M_{m}M_{n} \) and
\( M_{m} \) is the multiplicity of \( h_{m} \). We show that in
that case there exist a suitable norm to measure the regularity of
\( V \), denoted \( \epsilon _{V} \), and positive constants, \(
\epsilon _{\star } \)
and \( \delta _{\star } \), with the property: if \( \epsilon
_{V}<\epsilon _{\star } \)
then there exists a measurable subset \( \Omega _{\infty }\subset
\Omega _{0} \)
such that its Lebesgue measure fulfills \( |\Omega _{\infty }|\geq
|\Omega _{0}|-\delta _{\star }\epsilon _{V} \)
and the Floquet Hamiltonian has a pure point spectrum for all \(
\omega \in \Omega _{\infty } \).
\end{abstract}

\section{Introduction}

The problem we address in this paper concerns spectral analysis of
so called Floquet Hamiltonians. The study of stability of non
autonomous
quantum dynamical systems is an effective tool to understand most
of quantum problems which involve a small number of particles. When
these systems are time-periodic the spectral analysis of the evolution
operator over one period can give a fairly good information on this
stability, see e.g. \cite{EnssVeselic}. In fact this type of result
generalises the celebrated RAGE theorem concerned with
time-independent
systems (one can consult \cite{ReedSimon} for a summary). As shown
in \cite{Howland} and \cite{Yajima} the spectral analysis of the
evolution operator over one period (so called monodromy operator or
Floquet operator) is equivalent to the spectral analysis of the
corresponding
Floquet Hamiltonian (sometimes called operator of quasi-energy). This
is also what we are aiming for in this article. More precisely, we
analyse time-periodic quantum systems which are weakly regular in
time and \char`\"{}space\char`\"{} in the sense of an appropriately
chosen norm, and give sufficient conditions to insure that the Floquet
Hamiltonians has a pure point spectrum.

Such a program is not new. In the pioneering work \cite{Bellissard}
Bellissard has considered the so called pulsed rotor which is analytic
in time and space, using a KAM type algorithm. Then Combescure
\cite{Combescure}
was able to treat harmonic oscillators driven by sufficiently smooth
perturbations by adapting to quantum mechanics the well known
Nash-Moser
trick (c.f. \cite{Nash} and \cite{Moser}). Later on these ideas
have been extended to a wider class of systems in
\cite{DuclosStovicek};
it was even possible to require no regularity in space by using the
so called adiabatic regularisation, originally proposed in
\cite{Howland2}
and further extended in \cite{Nenciu}, \cite{Joye}. However none
of these papers can be considered as optimal in the sense of having
found the minimal value of regularity in time below which the Floquet
Hamiltonian ceases to be pure point.

Though it is impossible to mention all the relevant contributions
to the study of stability of time-dependent quantum systems we would
like to mention the following ones. Perturbation theory for a fixed
eigenvalue has been extended, in \cite{DuclosStovicekVittot}, to
Floquet Hamiltonians which generically have a dense point spectrum.
Bounded quasi-periodic time dependent perturbations of two level
systems
are considered in \cite{BleherJauslinLebowitz} whereas the case of
unbounded perturbation of one dimensional oscillators are studied
in \cite{BambusiGraffi}. Averaging methods combined with KAM
techniques
were described in \cite{Scherer} and \cite{JauslinGuerinThomas}.

In the present paper we attempt to further improve the KAM algorithm,
particularly having in mind more optimal assumptions as far as the
regularity in time is concerned. As a thorough analysis of the
algorithm
has shown this is possible owing to the fact that the algorithm
contains
several free parameters (for example the choice of norms in auxiliary
Banach spaces that are constructed during the algorithm) which may
be adjusted. This type of improvements is also illustrated on an
example
following Theorem \ref{MainTheorem} in Section \ref{Sec:MainTheorem}.
A more detailed discussion of this topic is postponed to concluding
remarks in Section \ref{Sec:ConcludingRemarks}.

Another generalisation is that in the present result (Theorem
\ref{MainTheorem})
we allow degenerate eigenvalues of the unperturbed Hamilton operator
(denoted \( H \) in what follows). The degeneracy of eigenvalues
\( h_{m} \) of \( H \) can grow arbitrarily fast with \( m \) provided
the time-dependent perturbation is sufficiently regular. To our
knowledge
this is a new feature in this context. Previously two conditions were
usually imposed, namely bounded degeneracy and a growing gap condition
on eigenvalues \( h_{m} \), reducing this way the scope of
applications
of this theory to one dimensional confined systems. Owing to the
generalisation
to degenerate eigenvalues we are able to consider also some models
in higher dimensions, for example the \( N \)-dimensional quantum
top, i.e., the \( N \)-dimensional version of the pulsed rotor. A
short description of this model is given, too, in Section
\ref{Sec:MainTheorem}
after Theorem \ref{MainTheorem}.

The article is organised as follows. In Section \ref{Sec:MainTheorem}
we introduce the notation and formulate the main theorem. The basic
idea of the KAM-type algorithm is outlined in Section \ref{Sec:
FormalLimitProc}.
The algorithm consists in an iterative procedure resulting in
diagonalisation
of the Floquet Hamiltonian. For this sake one constructs an auxiliary
sequence of Banach spaces which form in fact a directed sequence.
The procedure itself may formally be formulated in terms of an
inductive
limit. Sections
\ref{Sec:ConvergenceHilbertSpace}--\ref{Sec:ConstructSeqs}
contain some additional results needed for the proof, particularly
the details of the construction of the auxiliary Banach spaces and
how they are related to Hermitian operators in the given Hilbert
space,
and a construction of the set of \char`\"{}non-resonant\char`\"{}
frequencies for which the Floquet Hamiltonian has a pure point
spectrum
(the frequency is considered as a parameter). Section
\ref{Sec:ProofMainTheorem}
is devoted to the proof of Theorem \ref{MainTheorem}. In Section
\ref{Sec:ConcludingRemarks} we conclude our presentation with several
remarks concerning comparison of the result stated in Theorem
\ref{MainTheorem}
with some previous ones.

\section{\label{Sec:MainTheorem}Main theorem}

The central object we wish to study in this paper is a self-adjoint
operator of the form \( \bK +\bV  \) acting in the Hilbert space\[
\KK =L^{2}([\, 0,T\, ],\textrm{dt})\otimes \HH \cong L^{2}([\, 0,T\,
],\HH ,\textrm{dt})\]
where \( T=2\pi /\omega  \), \( \omega  \) is a positive number
(a frequency) and \( \HH  \) is a fixed separable Hilbert space.
The operator \( \bK  \) is self-adjoint and has the form\[
\bK =-i\, \partial _{t}\otimes 1+1\otimes H\]
where the differential operator \( -i\partial _{t} \) acts in \(
L^{2}([\, 0,T\, ],\textrm{dt}) \)
and represents the self-adjoint operator characterised by periodic
boundary conditions. This means that the eigenvalues of \( -i\partial
_{t} \)
are \( k\omega  \), \( k\in \Z  \), and the corresponding normalised
eigenvectors are \( \chi _{k}(t)=T^{-1/2}\exp (ik\omega t). \) \( H \)
is a self-adjoint operator in \( \HH  \) and is supposed to have
a discrete spectrum. Finally, \( \bV  \) is a bounded Hermitian
operator
in \( \KK  \) determined by a measurable operator-valued function
\( t\mapsto V(\omega t)\in \BB (\HH ) \) such that \( \sup _{t\in \R
}\| V(t)\| <\infty  \),
\( V(t) \) is \( 2\pi  \)-periodic, and for almost all \( t\in \R  \),
\( V(t)^{\ast }=V(t) \). Naturally, \( (\bV \psi )(t)=V(\omega t)\psi
(t) \)
in \( \KK \cong L^{2}([\, 0,T\, ],\HH ,\textrm{dt}) \).

Let\[
\sum _{k\in \Z }k\omega \, P_{k}\]
be the spectral decomposition of \( -i\partial _{t} \) in \(
L^{2}([\, 0,T\, ],\textrm{dt}) \)
and let\[
H=\sum _{m\in \N }h_{m}\, Q_{m}\]
be the spectral decomposition of \( H \) in \( \HH  \). Thus we
can write\[
\HH =\sideset {}{^{\oplus }}\sum _{m\in \N }\HH _{m}\]
where \( \HH _{m}=\Ran Q_{m} \) are the eigenspaces. We suppose that
the multiplicities are finite,\[
M_{m}=\dim \HH _{m}<\infty ,\textrm{ }\forall m\in \N .\]
Hence the spectrum of \( \bK  \) is pure point and its spectral
decomposition
reads\begin{equation}
\label{K_spec_decomp-def}
\bK =\sum _{k\in \Z }\sum _{m\in \N }(k\omega +h_{m})\, P_{k}\otimes
Q_{m},
\end{equation}
implying a decomposition of \( \KK  \) into a direct sum,\[
\KK =\sideset {}{^{\oplus }}\sum _{(k,m)\in \Z \times \N }\Ran
(P_{k}\otimes Q_{m})\, .\]

Here is some additional notation. Set\begin{equation}
\label{Vknm_eq_int0T}
V_{knm}=\frac{1}{T}\int ^{T}_{0}e^{-ik\omega t}\, Q_{n}V(\omega t)\,
Q_{m}\, \textrm{dt}=\frac{1}{2\pi }\int ^{2\pi }_{0}e^{-ikt}\,
Q_{n}V(t)\, Q_{m}\, \textrm{dt}\in \BB (\HH _{m},\HH _{n})\, .
\end{equation}
Further,\[
\Delta _{mn}=h_{m}-h_{n},\]
and\[
\Delta _{0}=\inf _{m\neq n}|\Delta _{mn}|.\]
Finally we set\[
\mu _{mn}=\left( \min \{M_{m},M_{n}\}\right) ^{1/2}M_{m}M_{n}.\]

Now we are able to formulate our main result. Though not indicated
explicitly in the notation the operator \( \bK +\bV  \) is considered
as depending on the parameter \( \omega  \).

\thmcount

\begin{thm}
\label{MainTheorem}Fix \( J>0 \) and set \( \Omega _{0}:=[\,
\frac{8}{9}J,\frac{9}{8}J\, ] \).
Assume that \( \Delta _{0}>0 \) and that there exists \( \sigma >0 \)
such that\[
\Delta _{\sigma }(J):=J^{\sigma }\sum _{\substack {m,n\in \N \cr
\Delta _{mn}>J/2}}\frac{\mu _{mn}}{(\Delta _{mn})^{\sigma }}<\infty
\, .\]

Then for every \( r>\sigma +\frac{1}{2} \) there exist positive
constants
(depending, as indicated, on \( \sigma  \), \( r \), \( \Delta _{0} \)
and \( J \) but independent of \( V \)), \( \epsilon _{\star
}(r,\Delta _{0},J) \)
and \( \delta _{\star }(\sigma ,r,J) \), with the property: if\[
\epsilon _{V}:=\ \sup _{n\in \N }\sum _{k\in \Z }\sum _{m\in \N }\|
V_{knm}\| \, \max \{|k|^{r},1\}<\min \left\{ \epsilon _{\star
}(r,\Delta _{0},J),\frac{|\Omega _{0}|}{\delta _{\star }(\sigma
,r,J)}\right\} \]
(here \( |\Omega _{\ast }| \) stands for the Lebesgue measure of
\( \Omega _{\ast } \)) then there exists a measurable subset \(
\Omega _{\infty }\subset \Omega _{0} \)
such that \begin{equation}
\label{measureJinfty_geq_RHS}
|\Omega _{\infty }|\geq |\Omega _{0}|-\delta _{\star }(\sigma ,r,J)\,
\epsilon _{V}
\end{equation}
and the operator \( \bK +\bV  \) has a pure point spectrum for all
\( \omega \in \Omega _{\infty } \)
\end{thm}
\begin{rems*}1) In the course of the proof we shall show even more.
Namely, for all \( \omega \in \Omega _{\infty } \) and any eigenvalue
of \( \bK +\bV  \) the corresponding eigen-projector \( P \) belongs
to the Banach algebra with the norm\[
\| P\| =\sup _{n\in \N }\sum _{k\in \Z }\sum _{m\in \N }\|
P_{knm}\| \, \max \{|k|^{r-\sigma -\frac{1}{2}},1\}.\]
This shows that \( P \) is \( (r-\sigma -1/2) \)-differentiable
as a map from \( [\, 0,T\, ] \) to the space of bounded operators
in \( \HH  \)

2) The constants \( \epsilon _{\star }(r,\Delta _{0},J) \) and \(
\delta _{\star }(\sigma ,r,J) \)
are in fact known quite explicitly and are given by formulae
(\ref{gammastar_bound1}),
(\ref{gammastar_bound2}), (\ref{formula_delta_1}) and
(\ref{formula_delta_2}).
Setting \( \alpha =2 \) and \( q^{r}=e^{2} \) in these formulae
(this is a possible choice) we get\[
\epsilon _{\star }(r,\Delta _{0},J)=\frac{\min \left\{4 \Delta
_{0},\, J\right\}}{270\, e^{3}} ,\]
and\begin{eqnarray*}
\delta _{\star }(\sigma ,r,J) & = & 1440\, e^{5}2^{\sigma }\left(
\frac{2\sigma +1}{\left( 1-e^{-\frac{2}{r}}\right) e}\right) ^{\sigma
+\frac{1}{2}}\left( \sum _{s=1}^{\infty
}s^{2}e^{-\frac{2}{r}(r-\sigma -\frac{1}{2})s}\right) \Delta _{\sigma
}(J)\\
 & = & 1440\left( \frac{2\sigma +1}{\left( 1-e^{-\frac{2}{r}}\right)
e}\right) ^{\sigma +\frac{1}{2}}2^{\sigma }e^{3+\frac{2}{r}\left(
\sigma +\frac{1}{2}\right) }\frac{1+e^{-2+\frac{2}{r}\left( \sigma
+\frac{1}{2}\right) }}{\left( 1-e^{-2+\frac{2}{r}\left( \sigma
+\frac{1}{2}\right) }\right) ^{3}}\, \Delta _{\sigma }(J)
\end{eqnarray*}

3) The formulae for \( \epsilon _{\star } \) and \( \delta _{\star }
\)
can be further simplified if we assume that \( r \) is not too big,
more precisely under the assumption that \( r\leq \frac{7}{8}\,
(2\sigma +1) \)
(if this is not the case we can always replace \( r \) by a smaller
value but still requiring that \( r>\sigma +\frac{1}{2} \)). A better
choice than that made in the previous remark is \( \alpha =2 \) and
\( q=e^{4/(2\sigma +1)} \). We get (c.f. (\ref{gammastar_bound2}))\[
\epsilon _{\star }(r,\Delta _{0},J)=\frac{\min \{4\, \Delta
_{0},J\}}{270\, e}\, e^{-4r/(2\sigma +1)}\geq \frac{\min \{4\, \Delta
_{0},J\}}{270\, e^{9/2}}\]

and (c.f. (\ref{formula_delta_1}) and (\ref{formula_delta_2}))\[
\delta _{\star }(\sigma ,r,J)=1440\, e\, 2^{\sigma }\left(
\frac{2\sigma +1}{\left( 1-e^{-\frac{4}{2\sigma +1}}\right) e}\right)
^{\sigma +\frac{1}{2}}e^{\frac{8r}{2\sigma +1}}\left( \sum
_{s=1}^{\infty }s^{2}e^{-2\frac{2r-2\sigma -1}{2\sigma +1}s}\right)
\Delta _{\sigma }(J)\, .\]
Using the estimate\[
\sum _{s=1}^{\infty }s^{2}e^{-2xs}=\frac{\cosh (x)}{4\, \sinh
(x)^{3}}\leq \frac{1}{4\, x^{3}}\]
we finally obtain\begin{eqnarray*}
\delta _{\star }(\sigma ,r,J) & \leq  & 45\, e\, 2^{\sigma }\left(
\frac{2\sigma +1}{\left( 1-e^{-\frac{4}{2\sigma +1}}\right) e}\right)
^{\sigma +\frac{1}{2}}e^{\frac{8r}{2\sigma +1}}\left( \frac{2\sigma
+1}{r-\sigma -\frac{1}{2}}\right) ^{3}\Delta _{\sigma }(J)\, .
\end{eqnarray*}

\end{rems*}

We conclude this section with a brief description of two models
illustrating
the effectiveness of Theorem \ref{MainTheorem}. In the first model
we set \( \HH =L^{2}([\, 0,1\, ],\textrm{dx}) \), \( H=-\partial
_{x}^{2} \)
with Dirichlet boundary conditions, and \( V(t)=z(t)x^{2} \) where
\( z(t) \) is a sufficiently regular \( 2\pi  \)-periodic function.
As shown in \cite{Seba} the spectral analysis of this simple model
is essentially equivalent to the analysis of the so called quantum
Fermi accelerator. The particularity of the latter model is that the
underlying Hilbert space itself is time-dependent, \( \HH
_{t}=L^{2}([\, 0,a(t)\, ],\textrm{dx}) \)
where \( a(t) \) is a strictly positive periodic function. The
time-dependent
Hamiltonian is \( -\partial _{x}^{2} \) with Dirichlet boundary
conditions.
Using a convenient transformation one can pass from the Fermi
accelerator
to the former model getting the function \( z(t) \) expressed in
terms of \( a(t) \), \( a'(t) \) and \( a''(t) \). But let us return
to the analysis of our model. Eigenvalues of \( H \) are
non-degenerate,
\( h_{m}=m^{2}\pi ^{2} \) for \( m\in \N  \), with normalised
eigenfunctions
equal to \( \sqrt{2}\sin (m\pi x) \). Note that in the notation we
are using in the present paper \( 0\notin \N  \). A straightforward
calculation gives\[
V_{knm}=z_{k}\times \left\{ \begin{array}{c}
\frac{8(-1)^{m+n}mn}{(m^{2}-n^{2})^{2}\pi ^{2}}\quad \textrm{if
}m\neq n,\\
\\
\frac{1}{3}-\frac{1}{2m^{2}\pi ^{2}}\quad \textrm{ if }m=n,
\end{array}\right. \]
where \( z_{k}=\frac{1}{2\pi }\int _{0}^{2\pi }e^{-ikt}z(t)\, dt \)
is the Fourier coefficient of \( z(t) \). Hence one derives that\[
\epsilon _{V}=\sup _{n\in \N }\left( \frac{1}{3}+\frac{2}{n^{2}\pi
^{2}}+\frac{4}{\pi ^{2}}\sum ^{n-1}_{j=1}\frac{1}{j^{2}}\right) \sum
_{k\in \Z }|z_{k}|\max \{|k|^{r},1\}=\sum _{k\in \Z }|z_{k}|\max
\{|k|^{r},1\}.\]
For any \( J>0 \), \( \Delta _{\sigma }(J) \) is finite if and only
if \( \sigma >1 \). On the other hand, to have \( \epsilon _{V} \)
finite it is sufficient that \( z(t)\in C^{s} \) where \(
s>r+1>\sigma +\frac{1}{2}+1>\frac{5}{2} \).
So \( z(t)\in C^{3} \) suffices for the theory to be applicable.
This may be compared to an older result in \cite{DuclosStovicek},
\S4.2, giving a much worse condition, namely \( z(t)\in C^{17} \).

The second model is the pulsed rotator in \( N \) dimensions. In
this case \( \HH =L^{2}(S^{N},d\mu ) \), with \( S^{N}\subset \R
^{N+1} \)
being the \( N \)-dimensional unit sphere with the standard
(rotationally
invariant) Riemann metric and the induced normalised measure \( d\mu
\),
and \( H=-\Delta _{LB} \) is the Laplace-Beltrami operator on \(
S^{N} \).
The spectrum of \( H \) is well known, \( \Spec
(H)=\{h_{m}\}_{m=0}^{\infty } \),
where\[
h_{m}=m(m+N-1)\]
and the multiplicities are\[
M_{m}={m+N\choose N}-{m+N-2\choose N}.\]
The time-dependent operator \( V(t) \) in \( \HH  \) acts via
multiplication,
\( (V(t)\varphi )(x)=v(t,x)\varphi (x) \), where \( v(t,x) \) is
a real measurable bounded function on \( \R \times S^{N} \) which
is \( 2\pi  \)-periodic in the variable \( t \). Consequently, \( \KK
\cong L^{2}([\, 0,T\, ]\times S^{N},dt\, d\mu ) \)
and \( (\bV \psi )(t,x)=v(\omega t,x)\psi (t,x) \). Note that the
asymptotic behaviour of the eigenvalues and the multiplicities, as
\( m\to \infty  \), is \( h_{m}\sim m^{2} \), \( M_{m}\sim
(2/(N-1)!)\, m^{N-1} \).
So \( \Delta _{\sigma }(J) \) is finite, for any \( J>0 \), if and
only if\[
\sum
_{m^{2}-n^{2}>J/2}\frac{n^{\frac{3}{2}(N-1)}m^{N-1}}{(m^{2}-n^{2})^{\sigma
}}<\infty \, .\]
To ensure this condition we require that \( \sigma
>\frac{5}{2}(N-1)+1 \).
Let us assume that there exist \( s,u\in \Z _{+} \) such that, for
any system of local (smooth) coordinates \( (y_{1},\ldots ,y_{N}) \)
on \( S^{N} \), the derivatives \( \partial _{t}^{\, \alpha }\partial
_{y_{1}}^{\, \beta _{1}}\ldots \partial _{y_{N}}^{\, \beta
_{N}}v(t,y_{1},\ldots ,y_{N}) \)
exist and are continuous for all \( \alpha  \), \( \beta  \), \(
\alpha \leq s \)
and \( \beta _{1}+\ldots +\beta _{N}\leq u \). If \( u\geq 4 \)
then \( [\, H,[\, H,V(t)\, ] \){]} is a well defined second order
differential operator with continuous coefficient functions and the
operator \( [\, H,[\, H,V(t)\, ]](1+H)^{-1} \) is bounded. Clearly,\[
\frac{(h_{m}-h_{n})^{2}}{1+h_{m}}\, Q_{n}V(t)Q_{m}=Q_{n}[\, H,[\,
H,V(t)\, ]](1+H)^{-1}Q_{m}.\]
Using this relation one derives an estimate on \( V_{knm} \),\[
\| V_{knm}\| \leq \textrm{const}\, \frac{1+\min
\{h_{n},h_{m}\}}{|k|^{s}(h_{m}-h_{n})^{2}}\, ,\]
valid for \( k\neq 0 \) and \( m\neq n \). The number\[
\sup _{n\in \Z _{+}}\sum _{m\in \Z _{+},\, m\neq n}\frac{1+\min
\{h_{n},h_{m}\}}{(h_{m}-h_{n})^{2}}\]
is finite. To see it one can employ the asymptotics of \( h_{m} \)
and the fact that the sequence\[
a_{n}=\sum _{m\in \Z _{+},\, m\neq n}\frac{1+\min
\{n^{2},m^{2}\}}{(m^{2}-n^{2})^{2}}=\left( 1+\frac{1}{n^{2}}\right)
\frac{\pi
^{2}}{12}-\frac{3}{16n^{2}}+\frac{5}{16n^{4}}-\frac{1}{2n}\sum
^{2n-1}_{m=1}\frac{1}{m}\, ,\]
\( n=1,2,3,\ldots  \), is bounded. It follows that the norm \(
\epsilon _{V} \)
is finite if \( s>r+1>\sigma
+\frac{1}{2}+1>\frac{5}{2}(N-1)+1+\frac{3}{2}=\frac{5}{2}\, N \).
Thus the theory is applicable provided \( u\geq 4 \) and \(
s>\frac{5}{2}N \).
The same example has also been treated by adiabatic methods in
\cite{Nenciu}.
In that case the assumptions are weaker. It suffices that \( v(t,x) \)
be \( (N+1) \)-times differentiable in \( t \) with all derivatives
\( \partial _{t}^{\, \alpha }v(t,x) \), \( 0\leq \alpha \leq N+1 \),
uniformly bounded. However the conclusion is somewhat weaker as well.
Under this assumption \( \bK +\bV  \) has no absolutely continuous
spectrum but nothing is claimed about the singular continuous
spectrum.

\section{\label{Sec: FormalLimitProc}Formal limit procedure}

Suppose there is given a directed sequence of real or complex Banach
spaces, \( \{\gX _{s}\}_{s=0}^{\infty } \), with linear mappings\[
\iota _{us}:\gX _{s}\to \gX _{u}\quad \textrm{if }s\leq u,\textrm{
with }\| \iota _{us}\| \leq 1,\]
(and \( \iota _{ss} \) is the unite mapping in \( \gX _{s} \)) and
such that\[
\iota _{vu}\iota _{us}=\iota _{vs}\quad \textrm{if }s\leq u\leq v\,
.\]
To simplify the notation we set in what follows\[
\iota _{s}=\iota _{s+1,s}.\]
Denote by \( \gX _{\infty } \) the norm inductive limit of \( \{\gX
_{s},\textrm{ }\iota _{us}\} \)
in the sense of \cite{Palmer}, \S1.3.4 or \cite{Sakai}, \S1.23 (the
algebraic inductive limit is endowed with a seminorm induced by \(
\limsup _{s}\| \cdot \| _{s} \),
the kernel of this seminorm is divided out and the result is
completed).
\( \gX _{\infty } \) is related to the original directed sequence
via the mappings \( \iota _{\infty s}:\gX _{s}\to \gX _{\infty } \)
obeying \( \| \iota _{\infty s}\| \leq 1 \) and \( \iota
_{\infty u}\iota _{us}=\iota _{\infty s} \)
if \( s\leq u \). By the construction, the union \( \bigcup _{s\geq
s_{0}}\iota _{\infty s}(\gX _{s}) \)
is dense in \( \gX _{\infty } \) for any \( s_{0}\in \Z _{+} \).

If \( \{A_{s}\in \BB (\gX _{s})\} \) is a family of bounded operators,
defined for \( s\geq s_{0} \) and such that\[
A_{u}\iota _{us}=\iota _{us}A_{s}\quad \textrm{if }s_{0}\leq s\leq
u,\textrm{ and }\sup _{s}\| A_{s}\| <\infty ,\]
then \( A_{\infty }\in \BB (\gX _{\infty }) \) designates the
inductive
limit of this family characterised by the property \( A_{\infty
}\iota _{\infty s}=\iota _{\infty s}A_{s} \),
\( \forall s\geq s_{0} \).

Let \( \DD _{\infty }\in \BB (\gX _{\infty }) \) be the inductive
limit of a family of bounded operators \( \{\DD _{s}\in \BB (\gX
_{s});\textrm{ }s\geq 0\} \),
with the property\begin{equation}
\label{normDsleq1}
\| \DD _{s}\| \leq 1,\textrm{ }\| 1-\DD _{s}\| \leq
1,\quad \forall s.
\end{equation}
We also suppose that there is given a sequence of one-dimensional
spaces \( \bk K_{s} \), \( s=0,1,\dots ,\infty  \), where the \(
K_{s} \)
are distinguished basis elements. Here the field \( \bk  \) is either
\( \C  \) or \( \R  \) depending on whether the Banach spaces \( \gX
_{s} \)
are complex or real. Set\[
\tilde{\gX }_{s}=\bk K_{s}\oplus \gX _{s},\quad s=0,1,\dots ,\infty
.\]
Then \( \{\tilde{\gX }_{s}\}_{s=0}^{\infty } \) becomes a directed
sequence of vector spaces provided one defines \( \tilde{\iota
}_{us}:\tilde{\gX }_{s}\to \tilde{\gX }_{u} \)
by\[
\tilde{\iota }_{us}|_{\gX _{s}}=\iota _{us}\textrm{ and }\tilde{\iota
}_{us}(K_{s})=K_{u}\quad \textrm{if }s\leq u.\]
Set\begin{equation}
\label{phi-def}
\phi (x)=\frac{1}{x}\left( e^{x}-\frac{e^{x}-1}{x}\right) =\sum
^{\infty }_{k=0}\frac{k+1}{(k+2)!}\, x^{k}\, .
\end{equation}
\propcount

\begin{prop}
\label{Prop:recforK+V}Suppose that, in addition to the sequences
\( \{\gX _{s}\}^{\infty }_{s=0} \), \( \{K_{s}\}_{s=0}^{\infty } \)
and \( \{\DD _{s}\}_{s=0}^{\infty } \), there are given sequences
\( \{V_{s}\}_{s=0}^{\infty } \) and \( \{\Theta
^{s}_{u}\}_{u=s+1}^{\infty } \)
such that \( V_{s}\in \gX _{s} \), \( \Theta ^{s}_{u}\in \BB (\gX
_{u}) \),
and\begin{equation}
\label{Thi_eq_iTh}
\Theta ^{s}_{v}\iota _{vu}=\iota _{vu}\Theta _{u}^{s}\quad \textrm{if
}s<u\leq v.
\end{equation}
Set\begin{equation}
\label{Ts_as_prod}
T_{s}=e^{\Theta ^{s-1}_{s}}e^{\Theta ^{s-2}_{s}}\ldots e^{\Theta
^{0}_{s}}\in \BB (\gX _{s})\quad \textrm{for }s\geq 1\, .
\end{equation}
Let \( \{W_{s}\}_{s=0}^{\infty } \) be another sequence, with \(
W_{s}\in \gX _{s} \),
defined recursively:\begin{eqnarray}
W_{0} & = & V_{0},\nonumber \\
W_{s+1} & = & \iota _{s}(W_{s})+T_{s+1}(V_{s+1}-\iota
_{s}(V_{s}))\label{recruleWs} \\
 &  & +\, \, \Theta ^{s}_{s+1}\phi (\Theta ^{s}_{s+1})\iota
_{s}(1-\DD _{s})(W_{s}-\iota _{s-1}(W_{s-1})),\nonumber
\end{eqnarray}
where we set, by convention, \( \gX _{-1}=0 \) , \( W_{-1}=0 \).
Extend the mappings \( \Theta ^{s}_{u} \) to \( \tilde{\Theta
}^{s}_{u}:\tilde{\gX }_{u}\to \tilde{\gX }_{u} \)
by\begin{equation}
\label{defcondTheta}
\tilde{\Theta }^{s}_{u}(K_{u})=-\Theta ^{s}_{u}\DD _{u}(\iota
_{us}(W_{s}))-(1-\DD _{u})(\iota _{us}(W_{s})-\iota
_{u,s-1}(W_{s-1})),
\end{equation}
and consequently the mappings \( T_{s} \) to \(
\tilde{T}_{s}:\tilde{\gX }_{s}\to \tilde{\gX }_{s} \),\[
\tilde{T}_{s}=e^{\tilde{\Theta }^{s-1}_{s}}e^{\tilde{\Theta
}^{s-2}_{s}}\ldots e^{\tilde{\Theta }^{0}_{s}}\textrm{ for }s\geq
1,\textrm{ }\tilde{T}_{0}=1.\]
Then it holds\begin{equation}
\label{T(K+V)}
\tilde{T}_{s}(K_{s}+V_{s})=K_{s}+\DD _{s}(W_{s})+(1-\DD
_{s})(W_{s}-\iota _{s-1}(W_{s-1})),\quad s=0,1,2,\dots .
\end{equation}

\end{prop}
\begin{rem*}
Since \( \tilde{\Theta }^{s}_{u}(K_{u})\in \gX _{u} \) it is easy
to observe that\[
\tilde{T}_{s}(K_{s})-K_{s}\in \gX _{s}.\]
Furthermore, note that (\ref{defcondTheta}) implies that \(
\tilde{\Theta }^{s}_{v}(K_{v})=\iota _{vu}\tilde{\Theta
}^{s}_{u}(K_{u}) \)
if \( 0\leq s<u\leq v \), and so the mappings \( \tilde{\Theta
}^{s}_{u} \)
still satisfy\[
\tilde{\Theta }^{s}_{v}\tilde{\iota }_{vu}=\tilde{\iota
}_{vu}\tilde{\Theta }_{u}^{s}\quad \textrm{if }s<u\leq v.\]

\end{rem*}
\begin{proof}
By induction in \( s \). For \( s=0 \) the claim is obvious. In
the induction step \( s\to s+1 \) one may use the induction hypothesis
and relations (\ref{defcondTheta}) and
(\ref{recruleWs}):\begin{eqnarray*}
\tilde{T}_{s+1}(K_{s+1}+V_{s+1}) & = & \tilde{T}_{s+1}\tilde{\iota
}_{s}(K_{s}+V_{s})+T_{s+1}(V_{s+1}-\iota _{s}(V_{s}))\\
 & = & e^{\tilde{\Theta }^{s}_{s+1}}\tilde{\iota
}_{s}\tilde{T}_{s}(K_{s}+V_{s})+T_{s+1}(V_{s+1}-\iota _{s}(V_{s}))\\
 & = & e^{\tilde{\Theta }^{s}_{s+1}}\tilde{\iota }_{s}(K_{s}+\DD
_{s}(W_{s})+(1-\DD _{s})(W_{s}-\iota _{s-1}(W_{s-1})))\\
 &  & +T_{s+1}(V_{s+1}-\iota _{s}(V_{s}))\\
 & = & K_{s+1}+\DD _{s+1}(\iota _{s}(W_{s}))+\frac{e^{\Theta
^{s}_{s+1}}-1}{\Theta ^{s}_{s+1}}\tilde{\Theta
}^{s}_{s+1}\tilde{\iota }_{s}(K_{s}+\DD _{s}(W_{s}))\\
 &  & +e^{\Theta ^{s}_{s+1}}\iota _{s}(1-\DD _{s})(W_{s}-\iota
_{s-1}(W_{s-1}))+T_{s+1}(V_{s+1}-\iota _{s}(V_{s}))\\
 & = & K_{s+1}-(1-\DD _{s+1})\iota _{s}(W_{s})+\iota
_{s}(W_{s})+T_{s+1}(V_{s+1}-\iota _{s}(V_{s}))\\
 &  & +\left( e^{\Theta ^{s}_{s+1}}-\frac{e^{\Theta
^{s}_{s+1}}-1}{\Theta ^{s}_{s+1}}\right) \iota _{s}(1-\DD
_{s})(W_{s}-\iota _{s-1}(W_{s-1}))\\
 & = & K_{s+1}-(1-\DD _{s+1})\iota _{s}(W_{s})+W_{s+1}\\
 & = & K_{s+1}+\DD _{s+1}(W_{s+1})+(1-\DD _{s+1})(W_{s+1}-\iota
_{s}(W_{s}))\, .
\end{eqnarray*}

\end{proof}
\propcount

\begin{prop}
\label{Prop:w_leq_dv}Assume that the sequences \(
\{V_{s}\}_{s=0}^{\infty } \),
\( \{W_{s}\}_{s=0}^{\infty } \) and \( \{\Theta
^{s}_{u}\}_{u=s}^{\infty } \)
have the same meaning and obey the same assumptions as in Proposition
\ref{Prop:recforK+V}. Denote\[
w_{s}=\| W_{s}-\iota _{s-1}(W_{s-1})\| \]
(with \( w_{0}=\| W_{0}\|  \)). Assume, in addition, that there
exist
a sequence of positive real numbers, \( \{F_{s}\}_{s=0}^{\infty } \),
such that\begin{equation}
\label{Theta_leq_Fw}
\| \Theta ^{s}_{u}\| \leq F_{s}w_{s},\quad \forall s,u,\textrm{
}u>s,
\end{equation}
a sequence of non-negative real numbers \( \{v_{s}\}^{\infty }_{s=0}
\)
such that\[
\| V_{s}-\iota _{s-1}(V_{s-1})\| \leq v_{s},\textrm{ }\forall
s,\]
(for \( s=0 \) this means \( \| V_{0}\| \leq v_{0} \)) and a
constant
\( A\geq 0 \) such that\begin{equation}
\label{Thmconv_assumA}
F_{s}v_{s}^{\, 2}\leq A\, v_{s+1},\textrm{ }\forall s,
\end{equation}
and that it holds true\begin{eqnarray}
B=\sum _{s=0}^{\infty }F_{s}v_{s}<\infty . & \label{Thmconv_assumB}
\end{eqnarray}
Denote\begin{equation}
\label{Thmconv_assumC}
C=\sup _{s}F_{s}v_{s}.
\end{equation}
If \( d>0 \) obeys\begin{equation}
\label{ABd_ineq}
e^{dB}+A\phi (dC)\, d^{2}\leq d
\end{equation}
then\begin{equation}
\label{ws_leq_dvs}
w_{s}\leq d\, v_{s},\textrm{ }\forall s.
\end{equation}

\end{prop}
\begin{proof}
We shall proceed by induction in \( s \). If \( s=0 \) then \(
v_{0}=w_{0}=\| V_{0}\|  \)
and (\ref{ws_leq_dvs}) holds true since (\ref{ABd_ineq}) implies
that \( d\geq 1 \). The induction step \( s\to s+1 \): according
to (\ref{recruleWs}), (\ref{Ts_as_prod}), (\ref{normDsleq1}) and
(\ref{ABd_ineq}), and owing to the fact that \( \phi (x) \) is
monotone,
we have\begin{eqnarray*}
w_{s+1} & \leq  & \| T_{s+1}\| \, v_{s+1}+\| \Theta
_{s+1}^{s}\| \, \phi (\| \Theta _{s+1}^{s}\| )\, w_{s}\\
 & \leq  & \exp \left( \sum ^{s}_{j=0}F_{j}w_{j}\right) v_{s+1}+\phi
(F_{s}w_{s})F_{s}w_{s}^{\, 2}\\
 & \leq  & \exp \left( d\sum ^{s}_{j=0}F_{j}v_{j}\right) v_{s+1}+\phi
(d\, F_{s}v_{s})F_{s}d^{2}v_{s}^{\, 2}\\
 & \leq  & e^{dB}v_{s+1}+\phi (dC)d^{2}A\, v_{s+1}\\
 & \leq  & d\, v_{s+1}.
\end{eqnarray*}

\end{proof}
\begin{rem*}
If\[
B\leq \frac{1}{3}\ln 2\quad \textrm{and}\quad A\, \phi (3C)\leq
\frac{1}{9}\]
then (\ref{ABd_ineq}) holds true with \( d=3 \).
\end{rem*}
Recall that \( \Theta ^{s}_{\infty }\in \BB (\gX _{\infty }) \) is
the unique bounded operator on \( \gX _{\infty } \) such that\[
\Theta _{\infty }^{s}\iota _{\infty u}=\iota _{\infty u}\Theta
_{u}^{s},\textrm{ }\forall u>s.\]
If (\ref{Theta_leq_Fw}) is true then its norm is estimated
by\begin{equation}
\label{Tinfty_leq_Fw}
\| \Theta ^{s}_{\infty }\| \leq F_{s}w_{s}.
\end{equation}
\corcount

\begin{cor}
\label{Cor:converg_VWT}Under the same assumptions as in Proposition
\ref{Prop:w_leq_dv}, if \( d>0 \) exists such that condition
(\ref{ABd_ineq})
is satisfied, and\begin{equation}
\label{Finf}
F_{\mathrm{inf}}=\inf _{s}F_{s}>0
\end{equation}
then the limits\[
V_{\infty }=\lim _{s\to \infty }\iota _{\infty s}(V_{s}),\quad
W_{\infty }=\lim _{s\to \infty }\iota _{\infty s}(W_{s})\]
exist in \( \gX _{\infty } \), the limit\[
T_{\infty }=\lim _{s\to \infty }e^{\Theta _{\infty }^{s-1}}\dots
e^{\Theta ^{0}_{\infty }}\]
exists in \( \BB (\gX _{\infty }) \), and \( T_{\infty }\in \BB (\gX
_{\infty }) \)
can be extended to a linear mapping \( \tilde{T}_{\infty }:\tilde{\gX
}_{\infty }\to \tilde{\gX }_{\infty } \)
by\begin{equation}
\label{T(K)infty}
\tilde{T}_{\infty }(K_{\infty })-K_{\infty }=\lim _{s\to \infty
}\iota _{\infty s}\left( \tilde{T}_{s}(K_{s})-K_{s}\right) ,
\end{equation}
with the limit existing in \( \gX _{\infty } \). These objects obey
the equality\begin{equation}
\label{TKVinfty}
\tilde{T}_{\infty }(K_{\infty }+V_{\infty })=K_{\infty }+\DD _{\infty
}(W_{\infty }).
\end{equation}

\end{cor}
\begin{proof}
If \( u\geq s \) then\begin{eqnarray*}
\| \iota _{\infty u}(V_{u})-\iota _{\infty s}(V_{s})\|  & = &
\big \| \sum _{j=s+1}^{u}\iota _{\infty j}(V_{j}-\iota
_{j-1}(V_{j-1}))\big \| \leq \sum _{j=s+1}^{u}v_{j}.
\end{eqnarray*}
Since\[
\sum _{s=0}^{\infty }v_{s}\leq \frac{1}{F_{\mathrm{inf}}}\sum
_{s=0}^{\infty }F_{s}v_{s}<\infty \]
the sequence \( \{\iota _{\infty s}(V_{s})\} \) is Cauchy in \( \gX
_{\infty } \)
and so \( V_{\infty }\in \gX _{\infty } \) exists. Under assumption
(\ref{ws_leq_dvs}) we can apply the same reasoning to the sequence
\( \{\iota _{\infty s}(W_{s})\} \) to conclude that the limit \(
W_{\infty }=\lim _{s\to \infty }\iota _{\infty s}(W_{s}) \)
exists in \( \gX _{\infty } \). Set\[
\bar{T}_{s}=e^{\Theta _{\infty }^{s-1}}\dots e^{\Theta ^{0}_{\infty
}}\quad \textrm{if }s\geq 1,\textrm{ and }\bar{T}_{0}=1.\]
If \( u\geq s \) then, owing to (\ref{Tinfty_leq_Fw}) and
(\ref{ws_leq_dvs}),
we have\begin{eqnarray*}
\| \bar{T}_{u}-\bar{T}_{s}\|  & \leq  & \left( \exp \left( \sum
^{u-1}_{j=s}\| \Theta ^{j}_{\infty }\| \right) -1\right) \exp
\left( \sum ^{s-1}_{j=0}\| \Theta ^{j}_{\infty }\| \right) \\
 & \leq  & \exp \left( d\sum ^{u-1}_{j=0}F_{j}v_{j}\right) -\exp
\left( d\sum ^{s-1}_{j=0}F_{j}v_{j}\right) .
\end{eqnarray*}
Assumption (\ref{Thmconv_assumB}) implies that \( \{\bar{T}_{s}\} \)
is a Cauchy sequence in \( \BB (\gX _{\infty }) \) and so \(
T_{\infty }\in \BB (\gX _{\infty }) \)
exists.

To show (\ref{T(K)infty}) let us first verify the
inequality\begin{equation}
\label{eThetaK-K}
\| e^{\tilde{\Theta }^{s}_{u}}(K_{u})-K_{u}\| \leq
\frac{1+dB}{F_{\textrm{inf}}}\left( e^{F_{s}w_{s}}-1\right) ,
\end{equation}
valid for all \( u>s \). Actually, using definition
(\ref{defcondTheta})
and assumption (\ref{Theta_leq_Fw}), we get\begin{eqnarray*}
\| e^{\tilde{\Theta }^{s}_{u}}(K_{u})-K_{u}\|  & \leq  &
\frac{e^{\| \Theta ^{s}_{u}\| }-1}{\| \Theta ^{s}_{u}\|
}\| \tilde{\Theta }^{s}_{u}(K_{u})\| \\
 & \leq  & \frac{e^{\| \Theta ^{s}_{u}\| }-1}{\| \Theta
^{s}_{u}\| }\left( \| \Theta ^{s}_{u}\| \| W_{s}\|
+\| W_{s}-\iota _{s-1}(W_{s-1})\| \right) \\
 & \leq  & \left( e^{F_{s}w_{s}}-1\right) \left( \| W_{s}\|
+\frac{1}{F_{s}}\right) .
\end{eqnarray*}
To finish the estimate note that (\ref{Thmconv_assumB}) and
(\ref{ws_leq_dvs})
imply\[
\| W_{s}\| =\sum ^{s}_{j=1}(\| W_{j}\| -\|
W_{j-1}\| )+\| W_{0}\| \leq \sum ^{\infty }_{j=0}dv_{j}\leq
\frac{d}{F_{\textrm{inf}}}\sum ^{\infty
}_{j=0}F_{j}v_{j}=\frac{dB}{F_{\textrm{inf}}}.\]
With the aid of an elementary identity,\[
a_{j}\dots a_{0}-1=a_{j}\dots a_{1}(a_{0}-1)+a_{j}\dots
a_{2}(a_{1}-1)+\dots +(a_{j}-1),\]
we can derive from (\ref{eThetaK-K}): if \( 0\leq s\leq t<u \)
then\begin{eqnarray*}
\| e^{\tilde{\Theta }^{t}_{u}}\dots e^{\tilde{\Theta
}^{s}_{u}}(K_{u})-K_{u}\|  & \leq  & e^{\| \Theta ^{t}_{u}\|
+\dots +\| \Theta ^{s+1}_{u}\| }\| e^{\tilde{\Theta
}^{s}_{u}}(K_{u})-K_{u}\| \\
 &  & +\, e^{\| \Theta ^{t}_{u}\| +\dots +\| \Theta
^{s+2}_{u}\| }\| e^{\tilde{\Theta
}^{s+1}_{u}}(K_{u})-K_{u}\| \\
 &  & +\dots +\| e^{\tilde{\Theta }^{t}_{u}}(K_{u})-K_{u}\| \\
 & \leq  & \frac{1+dB}{F_{\textrm{inf}}}\left( e^{F_{t}w_{t}+\dots
+F_{s+1}w_{s+1}}\left( e^{F_{s}w_{s}}-1\right) \right. \\
 &  & +\, e^{F_{t}w_{t}+\dots +F_{s+2}w_{s+2}}\left(
e^{F_{s+1}w_{s+1}}-1\right) \\
 &  & \left. +\dots +\left( e^{F_{t}w_{t}}-1\right) \right) \\
 & = & \frac{1+dB}{F_{\textrm{inf}}}\left( e^{F_{t}w_{t}+\dots
+F_{s}w_{s}}-1\right) .
\end{eqnarray*}
Set temporarily in this proof\[
\tau _{s}=\iota _{\infty s}(\tilde{T}_{s}(K_{s})-K_{s})\in \gX
_{\infty }.\]
If \( t\geq s \) then\begin{eqnarray*}
\tau _{t}-\tau _{s} & = & \iota _{\infty t}\left( e^{\tilde{\Theta
}^{t-1}_{t}}\dots e^{\tilde{\Theta }^{0}_{t}}(K_{t})-\iota
_{ts}e^{\tilde{\Theta }^{s-1}_{s}}\dots e^{\tilde{\Theta
}^{0}_{s}}(K_{s})\right) \\
 & = & \iota _{\infty t}\left( e^{\tilde{\Theta }^{t-1}_{t}}\dots
e^{\tilde{\Theta }^{0}_{t}}(K_{t})-e^{\tilde{\Theta }^{s-1}_{t}}\dots
e^{\tilde{\Theta }^{0}_{t}}(K_{t})\right) \\
 & = & \iota _{\infty t}\left( \left( e^{\Theta ^{t-1}_{t}}\dots
e^{\Theta ^{s}_{t}}-1\right) \left( e^{\tilde{\Theta
}^{s-1}_{t}}\dots e^{\tilde{\Theta }^{0}_{t}}(K_{t})-K_{t}\right)
\right. \\
 &  & \left. +\, e^{\tilde{\Theta }^{t-1}_{t}}\dots e^{\tilde{\Theta
}^{s}_{t}}(K_{t})-K_{t}\right) .
\end{eqnarray*}
Hence\begin{eqnarray*}
\| \tau _{t}-\tau _{s}\|  & \leq  &
\frac{1+dB}{F_{\textrm{inf}}}\left( \left( e^{F_{t-1}w_{t-1}+\dots
+F_{s}w_{s}}-1\right) \left( e^{F_{s-1}w_{s-1}+\dots
+F_{0}w_{0}}-1\right) \right. \\
 &  & \left. +\, e^{F_{t-1}w_{t-1}+\dots +F_{s}w_{s}}-1\right) \\
 & = & \frac{1+dB}{F_{\textrm{inf}}}\left( e^{F_{t-1}w_{t-1}+\dots
+F_{0}w_{0}}-e^{F_{s-1}w_{s-1}+\dots +F_{0}w_{0}}\right) .
\end{eqnarray*}
This shows that the sequence \( \{\tau _{s}\} \) is Cauchy and thus
the limit on the RHS of (\ref{T(K)infty}) exists.

We conclude that it holds true, in virtue of (\ref{T(K+V)}),
that\begin{eqnarray*}
\tilde{T}_{\infty }(K_{\infty }+V_{\infty }) & = & K_{\infty }+\lim
_{s\to \infty }\iota _{\infty s}(\tilde{T}_{s}(K_{s})-K_{s})+\lim
_{s\to \infty }\bar{T}_{s}\iota _{\infty s}(V_{s})\\
 & = & K_{\infty }+\lim _{s\to \infty }\iota _{\infty
s}(\tilde{T}_{s}(K_{s}+V_{s})-K_{s})\\
 & = & K_{\infty }+\lim _{s\to \infty }\iota _{\infty s}\big (\DD
_{s}(W_{s})+(1-\DD _{s})(W_{s}-\iota _{s-1}(W_{s-1}))\big )\\
 & = & K_{\infty }+\lim _{s\to \infty }\big (\DD _{\infty }(\iota
_{\infty s}(W_{s}))+(1-\DD _{\infty })(\iota _{\infty s}(W_{s})-\iota
_{\infty ,s-1}(W_{s-1}))\big )\\
 & = & K_{\infty }+\DD _{\infty }(W_{\infty }).
\end{eqnarray*}
So equality (\ref{TKVinfty}) has been verified as well.
\end{proof}

\section{\label{Sec:ConvergenceHilbertSpace}Convergence in a Hilbert
space}

Let \( \{\gX _{s},\iota _{us}\} \) be a directed sequence of real
or complex Banach spaces, as introduced in Section \ref{Sec:
FormalLimitProc}.
In this section it is sufficient to know that \( \KK  \) is a
separable
complex Hilbert space and \( \bK  \) is a closed (densely defined)
operator in \( \KK  \) . Suppose that for each \( s\in \Z _{+} \)
there is given a bounded linear mapping,\[
\kappa _{s}:\gX _{s}\to \BB (\KK ),\textrm{ with }\| \kappa
_{s}\| \leq 1,\]
and such that\[
\forall s,u,\textrm{ }0\leq s\leq u,\quad \kappa _{u}\iota
_{us}=\kappa _{s}.\]
If the Banach spaces \( \gX _{s} \) are real then the mappings \(
\kappa _{s} \)
are supposed to be linear over \( \R  \) otherwise they are linear
over \( \C  \). Then there exists a unique linear bounded mapping
\( \kappa _{\infty }:\gX _{\infty }\to \BB (\KK ) \) satisfying,
\( \forall s\in \Z _{+} \), \( \kappa _{\infty }\iota _{\infty
s}=\kappa _{s} \).
Clearly, \( \| \kappa _{\infty }\| \leq 1 \). Extend the
mappings
\( \kappa _{s} \) to \( \tilde{\kappa }_{s}:\tilde{\gX }_{s}=\bk
K_{s}+\gX _{s}\to \C \bK +\BB (\KK ) \)
by defining\[
\tilde{\kappa }_{s}(K_{s})=\bK ,\textrm{ }\forall s\in \Z _{+}\cup
\{\infty \}.\]
So \( \tilde{\kappa }_{s}(K_{s}+X)=\bK +\kappa _{s}(X) \), with \(
X\in \gX _{s} \),
is a closed operator in \( \KK  \) with \( \Dom (\bK +\kappa
_{s}(X))=\Dom (\bK ) \).

Suppose, in addition, that there exists \( \bD \in \BB (\BB (\KK )) \)
such that\[
\forall s\in \Z _{+},\quad \bD \kappa _{s}=\kappa _{s}\DD _{s}.\]
Then it holds true, \( \forall s\in \Z _{+} \), \( \forall X\in \gX
_{s} \),\[
\kappa _{\infty }\DD _{\infty }(\iota _{\infty s}X)=\kappa _{\infty
}\iota _{\infty s}\DD _{s}(X)=\kappa _{s}\DD _{s}(X)=\bD \kappa
_{s}(X)=\bD \kappa _{\infty }(\iota _{\infty s}X).\]
Since the set of vectors \( \{\iota _{\infty s}(X);\textrm{ }s\in \Z
_{+},\textrm{ }X\in \gX _{s}\} \)
is dense in \( \gX _{s} \) we get \( \kappa _{\infty }\DD _{\infty
}=\bD \kappa _{\infty } \).

\propcount

\begin{prop}
\label{Prop:converg_in_KK}Under the assumptions of Corollary
\ref{Cor:converg_VWT}
and those introduced above in this section, let \( \{\bA
_{s}\}^{\infty }_{s=0} \)
be a sequence of bounded operators in \( \KK  \) such
that,\begin{equation}
\label{kappaTheta_eq_commA}
\forall s,u,\textrm{ }0\leq s<u,\textrm{ }\forall X\in \gX _{u},\quad
\kappa _{u}\big (\Theta ^{s}_{u}(X)\big )=[\, \bA _{s},\kappa
_{u}(X)\, ],
\end{equation}
\vskip -6pt\[
\forall s\in \Z _{+},\quad \bA _{s}(\Dom \bK )\subset \Dom \bK ,\]
and\[
\forall s,u,\textrm{ }0\leq s<u,\quad [\, \bA _{s},\bK \, ]=\kappa
_{u}(\tilde{\Theta }^{s}_{u}(K_{u}))\big |_{\Dom (\bK )}.\]
Moreover, assume that\begin{equation}
\label{sum_As_less_infty}
\sum ^{\infty }_{s=0}\| \bA _{s}\| <\infty .
\end{equation}
Set\[
\bV =\kappa _{\infty }(V_{\infty }),\textrm{ }\bW =\kappa _{\infty
}(W_{\infty }).\]
Then the limit\begin{equation}
\label{U_eq_lim_expA}
\bU =\lim _{s\to \infty }e^{\bA _{s-1}}\ldots e^{\bA _{0}}
\end{equation}
exists in the operator norm, the element \( \bU \in \BB (\KK ) \)
has a bounded inverse, and it holds true that\[
\bU (\Dom \bK )=\Dom \bK \]
and\begin{equation}
\label{U(K_pl_V)Uinv_eq_K_pl_DW}
\bU (\bK +\bV )\bU ^{-1}=\bK +\bD (\bW ).
\end{equation}

\end{prop}
For the proof we shall need a lemma.

\lemcount

\begin{lem}
\label{Lem:exp-AKexpA}Assume that \( \HH  \) is a Hilbert space,
\( K \) is a closed operator in \( \HH  \), \( A,B\in \BB (\HH ) \),\[
A(\Dom K)\subset \Dom K,\]
and\[
[\, A,K\, ]=B\big |_{\Dom (K)}.\]
Then it holds, \( \forall \lambda \in \C  \),\begin{equation}
\label{elamdaA_DomK_eq_DomK}
e^{\lambda A}(\Dom K)=\Dom K
\end{equation}
and\[
e^{-\lambda A}Ke^{\lambda A}=K+\frac{e^{-\lambda \ad _{A}}-1}{\ad
_{A}}\, B.\]

\end{lem}
\begin{rem*}
Here and everywhere in what follows we use the standard notation:
\( \ad _{A}B=[\, A,B\, ] \) and so \( e^{\lambda \ad
_{A}}B=e^{\lambda A}B\, e^{-\lambda A} \).
\end{rem*}
\begin{proof}
Choose an arbitrary vector \( v\in \Dom (K) \) and set\[
\forall n\in \Z _{+},\quad v_{n}=\sum ^{n}_{k=0}\frac{\lambda
^{k}}{k!}\, A^{k}v\, .\]
Then \( v_{n}\in \Dom (K) \) and \( v_{n}\to e^{\lambda A}v \) as
\( n\to \infty  \). On the other hand,\begin{eqnarray*}
Kv_{n} & = & \sum ^{n}_{k=0}\frac{\lambda ^{k}}{k!}\, (K\,
A^{k}-A^{k}K)v+\sum ^{n}_{k=0}\frac{\lambda ^{k}}{k!}\, A^{k}Kv\\
 & = & -\sum ^{n}_{k=1}\frac{\lambda ^{k}}{k!}\, \sum
^{k-1}_{j=0}A^{j}B\, A^{k-1-j}v+\sum ^{n}_{k=0}\frac{\lambda
^{k}}{k!}\, A^{k}Kv\, .
\end{eqnarray*}
So the limit \( \lim _{n\to \infty }Kv_{n} \) exists. Consequently,
since \( K \) is closed, \( e^{\lambda A}(\Dom K)\subset \Dom K \).
But \( (e^{\lambda A})^{-1}=e^{-\lambda A} \) has the same property
and thus equality (\ref{elamdaA_DomK_eq_DomK}) follows. Furthermore,
the above computation also shows that\[
K\, e^{\lambda A}=-\sum ^{\infty }_{k=1}\frac{\lambda ^{k}}{k!}\,
\sum ^{k-1}_{j=0}A^{j}B\, A^{k-1-j}+e^{\lambda A}K\, .\]
Application of the following algebraic identity (easy to verify),\[
\sum ^{\infty }_{k=1}\frac{\lambda ^{k}}{k!}\, \sum
^{k-1}_{j=0}A^{j}B\, A^{k-1-j}=e^{\lambda A}\left(
\frac{1-e^{-\lambda \ad _{A}}}{\ad _{A}}\, B\right) ,\]
concludes the proof.
\end{proof}

\begin{proof}
[Proof of Proposition \ref{Prop:converg_in_KK}] We use notation of
Corollary \ref{Cor:converg_VWT}. From (\ref{kappaTheta_eq_commA})
follows that, \( \forall s,u,\textrm{ }0\leq s<u \), \( \forall X\in
\gX _{u} \),\[
\kappa _{\infty }\Theta ^{s}_{\infty }(\iota _{\infty u}X)=\kappa
_{u}\Theta ^{s}_{u}(X)=[\, \bA _{s},\kappa _{u}(X)\, ]=[\, \bA
_{s},\kappa _{\infty }(\iota _{\infty u}X)\, ].\]
Since the set of vectors \( \{\iota _{\infty u}(X);\textrm{
}s<u,\textrm{ }X\in \gX _{u}\} \)
is dense in \( \gX _{\infty } \), we get, \( \forall X\in \gX
_{\infty } \),
\( \kappa _{\infty }\Theta ^{s}_{\infty }(X)=[\, \bA _{s},\kappa
_{\infty }(X)\, ] \),
and hence\[
\kappa _{\infty }\left( e^{\Theta ^{s}_{\infty }}(X)\right) =e^{\bA
_{s}}\kappa _{\infty }(X)\, e^{-\bA _{s}}.\]

Set\[
\bU _{s}=e^{\bA _{s-1}}\ldots e^{\bA _{0}}\textrm{ for }s\geq
1,\textrm{ }\bU _{0}=1.\]
Assumption (\ref{sum_As_less_infty}) implies that both sequences
\( \{\bU _{s}\} \) and \( \{\bU _{s}^{\, -1}\} \) are Cauchy in
\( \BB (\KK ) \) and hence the limit (\ref{U_eq_lim_expA}) exists
in the operator norm, with \( \bU ^{-1}=\lim _{s\to \infty }\bU
_{s}^{\, -1}\in \BB (\KK ) \).
Moreover, \( \forall X\in \gX _{\infty } \),\begin{equation}
\label{kT(X)_eq_Uk(X)Uinv}
\kappa _{\infty }T_{\infty }(X)=\kappa _{\infty }\left( \lim _{s\to
\infty }e^{\Theta _{\infty }^{s-1}}\ldots e^{\Theta _{\infty
}^{0}}X\right) =\lim _{s\to \infty }\bU _{s}\kappa _{\infty }(X)\,
\bU _{s}^{\, -1}.
\end{equation}

Next let us compute \( \tilde{\kappa }_{s}\tilde{T}_{s}(K_{s}) \).
For \( 0\leq s<u \), set \( \bB _{s}=\kappa _{u}(\tilde{\Theta
}^{s}_{u}(K_{u}))\in \BB (\KK ) \).
\( \bB _{s} \) doesn't depend on \( u>s \) since if \( 0\leq s<u\leq
v \)
then\[
\kappa _{u}\big (\tilde{\Theta }^{s}_{u}(K_{u})\big )=\kappa _{v}\big
(\iota _{vu}\tilde{\Theta }^{s}_{u}(K_{u})\big )=\kappa _{v}\big
(\tilde{\Theta }^{s}_{v}(K_{v})\big ).\]
We can apply Lemma \ref{Lem:exp-AKexpA} to the operators \( \bK  \),
\( \bA _{s} \), \( \bB _{s} \) to conclude that \( e^{-\bA _{s}}(\Dom
\bK )=\Dom \bK  \)
and\begin{equation}
\label{expAKexp(-A)_eq_K_plus_bound}
e^{\bA _{s}}\bK \, e^{-\bA _{s}}=\bK +\frac{e^{\ad _{\bA
_{s}}}-1}{\ad _{\bA _{s}}}\, \bB _{s}.
\end{equation}
On the other hand,\[
\tilde{\kappa }_{u}\left( e^{\tilde{\Theta }^{s}_{u}}(K_{u})\right)
=\tilde{\kappa }_{u}\left( K_{u}+\frac{e^{\Theta ^{s}_{u}}-1}{\Theta
^{s}_{u}}\tilde{\Theta }_{u}^{s}(K_{u})\right) =\bK +\frac{e^{\ad
_{\bA _{s}}}-1}{\ad _{\bA _{s}}}\, \bB _{s}.\]
Thus \( \tilde{\kappa }_{u}\left( e^{\tilde{\Theta
}^{s}_{u}}(K_{u})\right) =e^{\bA _{s}}\bK \, e^{-\bA _{s}}. \)
Consequently, \( \bU _{s}(\Dom \bK )=\Dom \bK  \) and\begin{equation}
\label{tildekappas_tildeTs(Ks)}
\tilde{\kappa }_{s}\tilde{T}_{s}(K_{s})=\bU _{s}\bK \, \bU _{s}^{\,
-1}.
\end{equation}

Set \( \bC _{s}=\bU _{s}\bK \bU _{s}^{\, -1}-\bK  \). According to
(\ref{expAKexp(-A)_eq_K_plus_bound}), \( \bC _{s}\in \BB (\KK ) \).
Now we can compute, using relation (\ref{tildekappas_tildeTs(Ks)}),
a limit in \( \BB (\KK ) \),\begin{eqnarray*}
\bC  & = & \lim _{s\to \infty }\bC _{s}\, =\, \lim _{s\to \infty
}\kappa _{s}(\tilde{T}_{s}(K_{s})-K_{s})\\
 & = & \kappa _{\infty }\left( \lim _{s\to \infty }\iota _{\infty
s}(\tilde{T}_{s}(K_{s})-K_{s})\right) \\
 & = & \kappa _{\infty }(\tilde{T}_{\infty }(K_{\infty })-K_{\infty
}).
\end{eqnarray*}
So \( \bK +\bC =\tilde{\kappa }_{\infty }(\tilde{T}_{\infty
}(K_{\infty })) \).
From the closeness of \( \bK  \), the equality \( \bU _{s}\bK \bU
_{s}^{\, -1}=\bK +\bC _{s} \),
and from the fact that the sequences \( \{\bU _{s}^{\, \pm 1}\} \),
\( \{\bC _{s}\} \) converge one deduces that \( \bU ^{\pm 1}(\Dom \bK
)\subset \Dom \bK  \)
and hence, in fact, \( \bU ^{\pm 1}(\Dom \bK )=\Dom \bK  \). In
addition,\begin{equation}
\label{UKUinv_eq_kappaT(K)}
\bU \bK \bU ^{-1}=\bK +\bC =\tilde{\kappa }_{\infty
}\tilde{T}_{\infty }(K_{\infty }).
\end{equation}
Combining (\ref{kT(X)_eq_Uk(X)Uinv}) and (\ref{UKUinv_eq_kappaT(K)})
one finds that\[
\tilde{\kappa }_{\infty }\tilde{T}_{\infty }(X)=\bU \tilde{\kappa
}_{\infty }(X)\bU ^{-1},\textrm{ }\forall X\in \tilde{\gX }_{\infty
}.\]
To conclude the proof it suffices to apply the mapping \(
\tilde{\kappa }_{\infty } \)
to equality (\ref{TKVinfty}).
\end{proof}

\section{\label{Sec:ChoiceDirectSeqBanachSps}Choice of the directed
sequence
of Banach spaces}

Suppose that there are given a decreasing sequence of subsets of the
interval \( ]0,+\infty [ \), \( \Omega _{0}\supset \Omega _{1}\supset
\Omega _{2}\supset \ldots  \),
a decreasing sequence of positive real numbers \( \{\varphi
_{s}\}^{\infty }_{s=0} \)
and a strictly increasing sequence of positive real numbers \(
\{E_{s}\}^{\infty }_{s=0} \),
\( 1\leq E_{1}<E_{2}<\ldots  \).

We construct a complex Banach space \( {^{0}\gX }_{s} \), \( s\geq 0
\),
as a subspace\[
{^{0}\gX }_{s}\subset L^{\infty }\left( \Omega _{s}\times \Z \times
\N \times \N ,\, \sum _{n\in \N }\sideset {}{^{\oplus }}\sum _{m\in
\N }\BB (\HH _{m},\HH _{n})\right) \]
formed by those elements \( X=\{X_{knm}(\omega )\} \) which satisfy\[
X_{knm}(\omega )\in \BB (\HH _{m},\HH _{n}),\textrm{ }\forall \omega
\in \Omega _{s},\textrm{ }\forall (k,n,m)\in \Z \times \N \times \N
,\]
and have finite norm\begin{equation}
\label{norm_Xs-def}
\| X\| _{s}=\sup _{\substack {\omega ,\omega '\in \Omega
_{s}\cr \omega \neq \omega '}}\, \sup _{n\in \N }\, \sum _{k\in \Z
}\sum _{m\in \N }\left( \| X_{knm}(\omega )\| +\varphi _{s}\,
\| \partial X_{knm}(\omega ,\omega ')\| \right) e^{|k|/E_{s}}
\end{equation}
where the symbol \( \partial  \) designates the discrete derivative
in \( \omega  \),\[
\partial X(\omega ,\omega ')=\frac{X(\omega )-X(\omega ')}{\omega
-\omega '}\, .\]
 In fact, this norm is considered in Appendix B (c.f.
(\ref{norm_AppB-def})),
and it is shown there that \( {^{0}\gX }_{s} \) is an operator algebra
with respect to the multiplication rule (\ref{prodUV}).

Let \( \gX _{s}\subset {^{0}\gX }_{s} \) be a closed real subspace
formed by those elements \( X\in {^{0}\gX }_{s} \) which
satisfy,\begin{equation}
\label{X-Hermitian}
\forall (k,n,m)\in \Z \times \N \times \N ,\, \forall \omega \in
\Omega _{s},\quad X_{knm}(\omega )^{\ast }=X_{-k,m,n}(\omega )\in \BB
(\HH _{n},\HH _{m}).
\end{equation}
Note, however, that \( \gX _{s} \) is not an operator subalgebra
of \( {^{0}\gX }_{s} \).

The sequence of Banach spaces, \( \{\gX _{s}\}^{\infty }_{s=0} \),
becomes directed with respect to mappings of restriction in the
variable
\( \omega  \): if \( u\geq s \) then we set\[
\iota _{us}:\gX _{s}\to \gX _{u},\textrm{ }\iota _{us}(X)=X|_{\Omega
_{u}}.\]
Because of the monotonicity of the sequences \( \{\varphi _{s}\} \)
and \( \{E_{s}\} \) we clearly have \( \| \iota _{us}\| \leq 1
\).

Next we introduce a bounded operator \( \DD _{s}\in \BB (\gX _{s}) \)
as an operator which extracts the diagonal part of a
matrix,\begin{equation}
\label{D_s-def}
\DD _{s}(X)_{knm}(\omega )=\delta _{k0}\delta _{nm}X_{0nn}(\omega ).
\end{equation}
Clearly, \( \| \DD _{s}\| \leq 1 \) and \( \| 1-\DD
_{s}\| \leq 1 \).

Let\[
V\in L^{\infty }\left( \Z \times \N \times \N ,\, \sum _{n\in \N
}\sideset {}{^{\oplus }}\sum _{m\in \N }\BB (\HH _{m},\HH
_{n})\right) \]
be the element with the components \( V_{knm}\in \BB (\HH _{m},\HH
_{n}) \)
given in (\ref{Vknm_eq_int0T}). Since, by assumption, \( V(t) \)
is Hermitian for almost all \( t \) it hold true that\[
(V_{knm})^{\ast }=V_{-k,m,n}.\]
We still assume, as in Theorem \ref{MainTheorem}, that there exists
\( r>0 \) such that\begin{equation}
\label{epsilon_V-def}
\epsilon _{V}=\sup _{n\in \N }\sum _{k\in \Z }\sum _{m\in \N }\|
V_{knm}\| \, \max \{|k|^{r},1\}<\infty \, .
\end{equation}
Let us define elements \( V_{s}\in \gX _{s} \), \( s\geq 0 \),
by\begin{equation}
\label{cutoff_V}\begin{aligned}
(V_s)_{knm}(\omega) &\;=\; V_{knm}\phantom{0}\quad\mathrm{if\
}|k|<E_s\\
&\;=\; 0\phantom{V_{knm}}\quad\mathrm{if\ }|k|\geq E_s
\end{aligned}
\end{equation} For \( s\geq 1 \) we get an estimate,\begin{eqnarray}
\| V_{s}-\iota _{s-1}(V_{s-1})\| _{s} & = & \sup _{n\in \N
}\sum _{\substack {k\in \Z \cr E_{s-1}\leq |k|<E_{s}}}\sum _{m\in \N
}\| V_{knm}\| \, e^{|k|/E_{s}}\nonumber \\
 & \leq  & e\, \sup _{n\in \N }\sum _{k\in \Z }\sum _{m\in \N }\|
V_{knm}\| \, \frac{\max
\{|k|^{r},1\}}{(E_{s-1})^{r}}\label{estim_V-iV} \\
 & = & \frac{e\, \epsilon _{V}}{(E_{s-1})^{r}}\, .\nonumber
\end{eqnarray}
Similarly, for \( s=0 \), we get\[
\| V_{0}\| \leq e\, \epsilon _{V}.\]
It is convenient to set \( E_{-1}=1 \), \( V_{-1}=0 \).

The sequence \( \{K_{s}\}^{\infty }_{s=0} \) has the same meaning
as in Section \ref{Sec: FormalLimitProc}, i.e., each \( K_{s} \)
is a distinguished basis vector in a one-dimensional vector space
\( \R K_{s} \). Furthermore, a sequence \( \Theta ^{s}_{u}\in \BB
(\gX _{u}) \),
\( 0\leq s<u \), is supposed to satisfy rule (\ref{Thi_eq_iTh}).
Similarly as in Proposition \ref{Prop:recforK+V} we construct
sequences
\( T_{s}\in \BB (\gX _{s}) \), \( s\geq 1 \), and \( W_{s}\in \gX
_{s} \),
\( s\geq 0 \), using relations (\ref{Ts_as_prod}) and
(\ref{recruleWs}),
respectively.

\propcount

\begin{prop}
\label{Prop: A0B0C0_imply_Cor3}Suppose that it holds\begin{equation}
\label{hyp_estim_Theta}
\| \Theta ^{s}_{u}\| \leq \frac{5}{\varphi _{s+1}}\, \|
W_{s}-\iota _{s-1}(W_{s-1})\| _{s},\textrm{ }\forall s,u,\textrm{
}0\leq s<u,
\end{equation}
and set\begin{equation}
\label{A0b0C0-def}
A_{\star }=5e\, \sup _{s\geq 0}\frac{(E_{s})^{r}}{\varphi
_{s+1}(E_{s-1})^{2r}},\textrm{ }B_{\star }=5e\sum ^{\infty
}_{s=0}\frac{1}{\varphi _{s+1}(E_{s-1})^{r}},\textrm{ }C_{\star
}=5e\, \sup _{s\geq 0}\frac{1}{\varphi _{s+1}(E_{s-1})^{r}}\, .
\end{equation}
If\begin{equation}
\label{hyp_estim_A0B0C0}
\epsilon _{V}B_{\star }\leq \frac{1}{3}\ln 2\quad \textrm{and}\quad
\epsilon _{V}A_{\star }\phi (3\epsilon _{V}C_{\star })\leq \frac{1}{9}
\end{equation}
then the conclusions of Corollary 3 hold true, particularly, the
objects
\( V_{\infty },W_{\infty }\in \gX _{\infty } \), \( T_{\infty }\in
\BB (\gX _{\infty }) \)
and \( \tilde{T}_{\infty }\in \BB (\tilde{\gX }_{\infty }) \) exist
and satisfy the equality\[
\tilde{T}_{\infty }(K_{\infty }+V_{\infty })=K_{\infty }+\DD _{\infty
}(W_{\infty })\, .\]

\end{prop}
\begin{rem*}
Respecting estimates (\ref{estim_V-iV}) and (\ref{hyp_estim_Theta})
we set in what follows\begin{equation}
\label{F_v_defby_phi_eps_E}
F_{s}=\frac{5}{\varphi _{s+1}}\textrm{ and }v_{s}=\frac{e\, \epsilon
_{V}}{(E_{s-1})^{r}},\textrm{ }s\geq 0.
\end{equation}

\end{rem*}
\begin{proof}
Taking into account the defining relations (\ref{F_v_defby_phi_eps_E})
one finds that the constants \( A \), \( B \) and \( C \) introduced
in Proposition \ref{Prop:w_leq_dv} may be chosen as\begin{equation}
\label{ABC_defby_A0B0C0}
A=\epsilon _{V}A_{\star },\textrm{ }B=\epsilon _{V}B_{\star }\textrm{
and }C=\epsilon _{V}C_{\star }.
\end{equation}
The assumption (\ref{hyp_estim_A0B0C0}) implies that\begin{equation}
\label{B_leq_Aphi()_leq}
B\leq \frac{1}{3}\ln 2\quad \textrm{and}\quad A\, \phi (3C)\leq
\frac{1}{9}
\end{equation}
and so, according to the remark following Proposition
\ref{Prop:w_leq_dv},
inequality (\ref{ABd_ineq}) holds true with \( d=3 \). Since \(
F_{\inf }=5/\varphi _{1}>0 \)
assumption (\ref{Finf}) of Corollary \ref{Cor:converg_VWT} as well
as all assumptions of Proposition \ref{Prop:w_leq_dv} are satisfied
and so the conclusions of Corollary \ref{Cor:converg_VWT} hold true.
\end{proof}

\section{Relation of the Banach spaces \protect\( \gX _{s}\protect \)
to
Hermitian operators in \protect\( \KK \protect \)}

The real Banach spaces \( \gX _{s} \) have been chosen in the previous
section. Set\[
\Omega _{\infty }=\bigcap ^{\infty }_{s=0}\Omega _{s}.\]
Suppose that \( \Omega _{\infty }\neq \emptyset  \) and fix \( \omega
\in \Omega _{\infty } \)
(so \( \omega >0 \)).

To an operator-valued function \( [\, 0,T\, ]\ni t\mapsto X(t)\in \BB
(\HH ) \)
there is naturally related an operator \( \bX  \) in \( \KK
=L^{2}([\, 0,T\, ],\HH ,\textrm{dt}) \)
defined by \( (\bX \psi )(t)=X(t)\psi (t) \). As is well known,\[
\| \bX \| \leq \| X\| _{SH}\]
where \( \| \cdot \| _{SH} \) is the so called Schur-Holmgren
norm,\begin{eqnarray}
\| X\| _{SH} & = & \max \left\{ \sup _{(\ell ,n)\in \Z \times
\N }\sum _{(k,m)\in \Z \times \N }\big \| P_{\ell }\otimes
Q_{n}\bX P_{k}\otimes Q_{m}\big \| \right. ,\nonumber \\
 &  & \phantom {max\{\{}\left. \sup _{(k,m)\in \Z \times \N }\sum
_{(\ell ,n)\in \Z \times \N }\big \| P_{\ell }\otimes Q_{n}\bX
P_{k}\otimes Q_{m}\big \| \right\} \label{norm_bX_leq_SH} \\
 & = & \max \left\{ \sup _{n\in \N }\sum _{k\in \Z }\sum _{m\in \N
}\| X_{knm}\| ,\, \sup _{m\in \N }\sum _{k\in \Z }\sum _{n\in
\N }\| X_{knm}\| \right\} .\nonumber
\end{eqnarray}
Here\[
X_{knm}=\frac{1}{T}\int ^{T}_{0}e^{-i\omega kt}\, Q_{n}X(t)Q_{m}\,
\textrm{dt}\, .\]
It is also elementary to verify that the Schur-Holmgren norm is an
operator norm, \( \| XY\| _{SH}\leq \| X\| _{SH}\|
Y\| _{SH} \),
with respect to the multiplication rule (\ref{prodUV}).

If \( X(t) \) is Hermitian for (almost) every \( t\in [\, 0,T\, ] \)
then it holds, \( \forall (k,n,m) \), \( (X_{knm})^{\ast }=X_{-k,m,n}
\),
and so\[
\| X\| _{SH}=\sup _{n\in \N }\sum _{k\in \Z }\sum _{m\in \N
}\| X_{knm}\| \, .\]
 Note also that, \( \forall s\in \Z _{+} \), \( \forall X\in \gX _{s}
\),\[
\| X(\omega )\| _{SH}\leq \| X\| _{s}\]
and, consequently, the same is also true for \( s=\infty  \).

To an element \( X\in {^{0}\gX _{s}}\subset L^{\infty }\left( \Omega
_{s}\times \Z \times \N \times \N ,\, \sideset {}{}\sum _{n\in \N
}\sideset {}{^{\oplus }}\sum _{\! \! \! \! \! m\in \N }\BB (\HH
_{m},\HH _{n})\right)  \)
such that \( \| X(\omega )\| _{SH}<\infty  \) we can relate an
operator-valued
function defined on the interval \( [\, 0,T\, ] \),\[
t\mapsto \sum _{k\in \Z }\sum _{n\in \N }\sum _{m\in \N }e^{ik\omega
t}\, X_{knm}(\omega )\, .\]
The corresponding operator in \( \KK  \) is denoted by \( \kappa
_{s}(X) \),
with a norm being bounded from above by \( \| X(\omega )\|
_{SH} \).
In particular, \( \forall X\in \gX _{s} \),\[
\| \kappa _{s}(X)\| \leq \| X(\omega )\| _{SH}\leq \|
X\| _{s}.\]
In addition, if \( X\in \gX _{s} \) then the operator \( \kappa
_{s}(X) \)
is Hermitian due to the property (\ref{X-Hermitian}) of \( X \).
This way we have introduced the mappings \( \kappa _{s}:\gX _{s}\to
\BB (\KK ) \)
for \( s\in \Z _{+} \).

Another property we shall need is that \( \kappa _{s} \) is an algebra
morphism in the sense: if \( X,Y\in {^{0}\gX _{s}} \) such that \(
\| X(\omega )\| _{SH}<\infty  \)
and \( \| Y(\omega )\| _{SH}<\infty  \) then \( \|
(XY)(\omega )\| _{SH}<\infty  \)
and\[
\kappa _{s}(XY)=\kappa _{s}(X)\kappa _{s}(Y)\, .\]
Particularly this is true for all \( X,Y\in \gX _{s} \).

Let \( \bD \in \BB (\BB (\KK )) \) be the operator on \( \BB (\KK ) \)
taking the diagonal part of an operator \( X\in \BB (\KK ) \),\[
\bD (X)=\sum _{k\in \Z }\sum _{m\in \N }P_{k}\otimes Q_{m}X\,
P_{k}\otimes Q_{m}.\]
Clearly, \( \bD \kappa _{s}=\kappa _{s}\DD _{s} \). Since\[
\| \bD (X)\| =\sup _{(k,m)\in \Z \times \N }\| P_{k}\otimes
Q_{m}X\, P_{k}\otimes Q_{m}\| \leq \| X\| \]
we have \( \| \bD \| \leq 1 \).

A consequence of (\ref{epsilon_V-def}) is that \( V=\{V_{knm}\} \)
has a a finite Schur-Holmgren norm, \( \| V\| _{SH}<\infty  \).
Let
\( V_{s}\in \gX _{s} \), \( s\in \Z _{+} \), be the cut-offs of
\( V \) defined in (\ref{cutoff_V}). Then\begin{eqnarray*}
\| V-V_{s}\| _{SH} & = & \sup _{n\in \N }\sum _{k\in \Z ,\,
|k|\geq E_{s}}\sum _{m\in \N }\| V_{knm}\| \\
 & \leq  & \frac{1}{(E_{s})^{r}}\, \sup _{n\in \N }\sum _{k\in \Z
}\sum _{m\in \N }\| V_{knm}\| \, \max \{|k|^{r},1\}\\
 & = & \frac{\epsilon _{V}}{(E_{s})^{r}}\, .
\end{eqnarray*}
We shall impose an additional condition on the increasing sequence
\( \{E_{s}\} \) of positive real numbers that occur in the definition
of the norm \( \| \cdot \| _{s} \) in \( \gX _{s} \) (c.f.
(\ref{norm_Xs-def})),
namely we shall require\begin{equation}
\label{limEs_eq_infty}
\lim _{s\to \infty }E_{s}=+\infty \, .
\end{equation}
In this case \( \lim _{s\to \infty }\| V-V_{s}\| _{SH}=0 \) and
so\begin{equation}
\label{V_eq_lim_kappa(V)-opnorm}
\bV =\lim _{s\to \infty }\kappa _{s}(V_{s})\quad \textrm{in the
operator norm}.
\end{equation}

We also assume that there exist \( A_{s}\in \gX _{s+1} \), \( s\in \Z
_{+} \),
such that\begin{equation}
\label{As_anti-Hermitian}
(A_{s})_{knm}(\omega )^{\ast }=-(A_{s})_{-k,m,n}(\omega ),
\end{equation}
and, using these elements, we define mappings \( {^{0}\Theta
}_{u}^{s}\in \BB ({^{0}\gX }_{u}) \),
\( u>s \), by\begin{equation}
\label{0Theta_eq_comm_As_X}
{^{0}\Theta }_{u}^{s}(X)=[\, \iota _{u,s+1}(A_{s}),X\, ]
\end{equation}
(where the commutator on the RHS makes sense since \( {^{0}\gX }_{u}
\)
is an operator algebra). Clearly, \( \| {^{0}\Theta }_{u}^{s}\|
\leq 2\| A_{s}\| _{s+1} \).
One finds readily that \( \gX _{u}\subset {^{0}\gX }_{u} \) is an
invariant subspace with respect to the mapping \( {^{0}\Theta
}_{u}^{s} \)
and so one may define \( \Theta _{u}^{s}={^{0}\Theta }_{u}^{s}\big
|_{\gX _{u}}\in \BB (\gX _{u}) \).
Since \( iA_{s}\in \gX _{s+1} \) we can set\[
\bA _{s}=-i\, \kappa _{s+1}(iA_{s+1})\in \BB (\KK ).\]
Clearly, \( \bA _{s} \) is anti-Hermitian and satisfies \( \| \bA
_{s}\| \leq \| A_{s}\| _{s+1} \).
Note that (\ref{0Theta_eq_comm_As_X}) implies that, \( \forall
s,u,\textrm{ }0\leq s<u \),
\( \forall X\in \gX _{u} \),\[
\kappa _{u}\big (\Theta _{u}^{s}(X)\big )=[\, \bA _{s},\kappa
_{u}(X)\, ]\, .\]

\lemcount

\begin{lem}
\label{Lem: comm_A_K}Let \( \{W_{s}\}^{\infty }_{s=0} \) be a sequence
of elements \( W_{s}\in \gX _{s} \) and let \( \tilde{\Theta
}_{u}^{s}:\tilde{\gX }_{u}\to \tilde{\gX }_{u} \)
be the extension of \( \Theta _{u}^{s} \), \( 0\leq s<u \), defined
in (\ref{defcondTheta}). Assume that the elements \( A_{s}\in
{^{0}\gX }_{s+1} \),
\( s\in \Z _{+} \), satisfy\begin{eqnarray}
 &  & (k\omega -\Delta _{mn})\, (A_{s})_{knm}(\omega
)\label{komega-Delta_As_eq_RHS} \\
 &  & \qquad =\left( \Theta _{u}^{s}\big (\iota _{us}\DD
_{s}(W_{s})\big )+\iota _{us}(1-\DD _{s})\big (W_{s}-\iota
_{s-1}(W_{s-1})\big )\right) _{knm}(\omega ),\nonumber
\end{eqnarray}
\( \forall (k,m,n)\in \Z \times \N \times \N  \), \( \forall
s,u,\textrm{ }0\leq s<u \).

Then it holds true that,\[
\forall s\in \Z _{+},\quad \bA _{s}(\Dom \bK )\subset \Dom \bK ,\]
and\[
\forall s,u,\textrm{ }0\leq s<u,\quad [\, \bA _{s},\bK \, ]=\kappa
_{u}(\tilde{\Theta }^{s}_{u}(K_{u}))\big |_{\Dom (\bK )}.\]

\end{lem}
\begin{proof}
Set \[
\bB _{s}=-\kappa _{u}\big (\tilde{\Theta }_{u}^{s}(K_{u})\big ).\]
Since the RHS of (\ref{komega-Delta_As_eq_RHS}) is in fact a matrix
entry of \( -\tilde{\Theta }_{u}^{s}(K_{u}) \) (c.f.
(\ref{defcondTheta}))
this assumption may be rewritten as the equality\[
\bK \, P_{\ell }\otimes Q_{n}\bA _{s}P_{k}\otimes Q_{m}=P_{\ell
}\otimes Q_{n}\bA _{s}P_{k}\otimes Q_{m}\bK +P_{\ell }\otimes
Q_{n}\bB _{s}P_{k}\otimes Q_{m},\]
valid for all \( (\ell ,n),(k,m)\in \Z \times \N  \). Since \( \bK  \)
is closed one easily derives from the last property that it holds
true, \( \forall (k,m)\in \Z \times \N  \),\begin{equation}
\label{KAsPk_otimes_Qm_eq_RHS}
\bK \, \bA _{s}P_{k}\otimes Q_{m}=\bA _{s}P_{k}\otimes Q_{m}\bK +\bB
_{s}P_{k}\otimes Q_{m}.
\end{equation}
Particularly, \( \bA _{s}\Ran (P_{k}\otimes Q_{m})\subset \Dom (\bK )
\).
But \( \Ran (P_{k}\otimes Q_{m}) \) are mutually orthogonal
eigenspaces
of \( \bK  \). Consequently, if \( v\in \Dom (\bK ) \), then the
sequence \( \{v_{N}\}^{\infty }_{N=1} \),\[
v_{N}=\sum _{k,\, |k|\leq N}\sum _{m,\, m\leq N}P_{k}\otimes Q_{m}v\]
has the property: \( v_{N}\to v \) and \( \bK v_{N}\to \bK v \),
as \( N\to \infty  \). Equality (\ref{KAsPk_otimes_Qm_eq_RHS}) implies
that\[
\bK \bA _{s}v_{N}=\bA _{s}\bK v_{N}+\bB _{s}v_{N},\textrm{ }\forall
N.\]
Again owing to the fact that \( \bK  \) is closed one concludes that
\( \bA _{s}v\in \Dom (\bK ) \) and \( \bK \bA _{s}v=\bA _{s}\bK v+\bB
_{s}v \).
\end{proof}
\propcount

\begin{prop}
\label{Prop:diag_of_K_p_V}Assume that \( \omega \in \Omega _{\infty }
\)
and the norms \( \| \cdot \| _{s} \) in the Banach spaces \(
\gX _{s} \)
satisfy (\ref{limEs_eq_infty}). Let \( \Theta _{u}^{s}\in \BB (\bX
_{u}) \),
\( 0\leq s<u \), be the operators defined in
(\ref{0Theta_eq_comm_As_X})
with the aid of elements \( A_{s}\in {^{0}\gX }_{s+1} \) satisfying
(\ref{As_anti-Hermitian}), and let \( W_{s}\in \gX _{s} \), \( s\in
\Z _{+} \),
be a sequence defined recursively in accordance with
(\ref{recruleWs}).
Assume that the elements \( A_{s} \), \( s\in \Z _{+} \), satisfy
condition (\ref{komega-Delta_As_eq_RHS}) and that\begin{equation}
\label{normAs_leq_RHS}
\| A_{s}\| \leq \frac{5}{2\varphi _{s+1}}\, \| W_{s}-\iota
_{s-1}(W_{s-1})\| ,\textrm{ }\forall s\in \Z _{+}.
\end{equation}
Moreover, assume that the numbers \( A_{\star } \), \( B_{\star } \),
\( C_{\star } \), as defined in (\ref{A0b0C0-def}), satisfy condition
(\ref{hyp_estim_A0B0C0}).

Then there exist, in \( \KK  \), a unitary operator \( \bU  \) and
a bounded Hermitian operator \( \bW  \) such that\[
\bU (\Dom \bK )=\Dom \bK \]
and\[
\bU (\bK +\bV )\bU ^{-1}=\bK +\bD (\bW ).\]

\end{prop}
\begin{proof}
The norm of \( \Theta ^{s}_{u} \) may be estimated as\[
\| \Theta ^{s}_{u}\| \leq 2\| A_{s}\| \leq
\frac{5}{\varphi _{s+1}}\, \| W_{s}-\iota _{s-1}(W_{s-1})\| .\]
This way the assumptions of Proposition \ref{Prop: A0B0C0_imply_Cor3}
are satisfied and consequently, according to Proposition \ref{Prop:
A0B0C0_imply_Cor3}
(and its proof), the same is true for Proposition \ref{Prop:w_leq_dv}
and Corollary \ref{Cor:converg_VWT} (with \( F_{s} \) and \( v_{s} \)
defined in (\ref{F_v_defby_phi_eps_E}) and the constants \( A \),
\( B \), \( C \) defined in (\ref{ABC_defby_A0B0C0})). Since it
holds \( \| \bA _{s}\| \leq \| A_{s}\| \leq
\frac{1}{2}F_{s}w_{s} \)
(where \( F_{s}=5/\varphi _{s+1} \)) and, by assumption, condition
(\ref{ABd_ineq}) is satisfied with \( d=3 \) we get\[
\sum _{s=0}^{\infty }\| \bA _{s}\| \leq \frac{1}{2}\sum
_{s=0}^{\infty }F_{s}w_{s}\leq \frac{3}{2}\sum _{s=0}^{\infty
}F_{s}v_{s}=\frac{3B}{2}<\infty .\]
This verifies assumption (\ref{sum_As_less_infty}) of Proposition
\ref{Prop:converg_in_KK}; the other assumptions of this proposition
are verified as well as follows from Lemma \ref{Lem: comm_A_K}. Note
that, in virtue of (\ref{V_eq_lim_kappa(V)-opnorm}), \( \kappa
_{\infty }(V_{\infty })=\lim _{s\to \infty }\kappa _{s}(V_{s}) \)
coincides with the given operator \( \bV  \). Furthermore, \( \bW
=\kappa _{\infty }(W_{\infty })=\lim _{s\to \infty }\kappa
_{s}(W_{s}) \)
is a limit of Hermitian operators and so is itself Hermitian, and
\( \bU =\lim _{s\to \infty }e^{\bA _{s-1}}\ldots e^{\bA _{0}} \)
is unitary. Equality (\ref{U(K_pl_V)Uinv_eq_K_pl_DW}) holds true
and this concludes the proof.
\end{proof}

\section{Set of non-resonant frequencies}

Let \( J>0 \) be fixed and assume that, \( \forall s\in \Z _{+} \),\[
\Omega _{s}\subset \big [\, \frac{8}{9}J,\frac{9}{8}J\, \big ].\]

The following definition concerns indices \( (k,n,m) \) corresponding
to non-diagonal entries, i.e., those indices for which either \(
k\neq 0 \)
or \( m\neq n \). The diagonal indices, with \( k=0 \) and \( m=n \),
will always be treated separately and, in fact, in a quite trivial
manner.

\begin{defn*}
We shall say that a multi-index \( (k,n,m)\in \Z \times \N \times \N
\)
is \emph{critical} if \( m\neq n \) and\begin{equation}
\label{def:crit_indcs}
\frac{kJ}{\Delta _{mn}}\in \, \big ]\frac{1}{2},2\big [
\end{equation}
(hence \( \sgn (k)=\sgn (h_{m}-h_{n})\neq 0 \)). In the opposite
case the multi-index will be called \emph{non-critical}.
\end{defn*}
\vskip 0pt

\begin{defn*}
Let \( \psi (k,n,m) \) be a positive function defined on non-diagonal
indices and \( W\in \gX _{s} \). A frequency \( \omega \in \Omega
_{s} \)
will be called \( (W,\psi ) \)--\emph{non-resonant} if for all
non-diagonal
indices \( (k,n,m)\in \Z \times \N \times \N  \) it
holds\begin{equation}
\label{distSpecSpec_geq_psi}
\dist \left( \Spec (k\omega -\Delta _{mn}+W_{0nn}(\omega )),\, \Spec
(W_{0mm}(\omega ))\right) \geq \psi (k,n,m).
\end{equation}
In the opposite case \( \omega  \) will be called \( (W,\psi )
\)--\emph{resonant}.
\end{defn*}
Note that, in virtue of (\ref{X-Hermitian}), \( W_{0mm}(\omega ) \)
is a Hermitian operator in \( \HH _{m} \).

\lemcount

\begin{lem}
\label{Lem:estim_Omegabad_s}Assume that \( \Omega _{s}\subset [\,
\frac{8}{9}J,\frac{9}{8}J\, ] \),
\( W\in \gX _{s} \) and \( \psi  \) is a positive function defined
on non-diagonal indices and obeying a symmetry
condition,\begin{equation}
\label{psi-symmetry-cond}
\psi (-k,m,n)=\psi (k,n,m)\ \textrm{ for all }(k,n,m)\
\textrm{non-diagonal} .
\end{equation}

If\begin{equation}
\label{part_W_0mm_leq_1_over_4}
\forall m\in \N ,\, \textrm{ }\forall \omega ,\omega '\in \Omega
_{s},\textrm{ }\omega \neq \omega ',\textrm{ }\| \partial
W_{0mm}(\omega ,\omega ')\| \leq \frac{1}{4},
\end{equation}
and if condition (\ref{distSpecSpec_geq_psi}) is satisfied for all
\( \omega \in \Omega _{s} \) and all non-critical indices \( (k,n,m)
\)
then the Lebesgue measure of the set \( \Omegabad _{s}\subset \Omega
_{s} \)
formed by \( (W,\psi ) \)--resonant frequencies may be estimated
as \emph{\begin{equation}
\label{Omegabads_leq_sums_psi}
|\Omegabad _{s}|\leq 8\sum _{\substack {m,n\in \N ,\cr \Delta _{mn}\,
>\, \frac{1}{2}J}}\, \sum _{\substack {k\in \N ,\cr \frac{\Delta
_{mn}}{2J}<k<\frac{2\Delta _{mn}}{J}}}\frac{M_{m}M_{n}}{k}\, \psi
(k,n,m)\, .
\end{equation}
}
\end{lem}
\begin{proof}
Let \( \lambda _{1}^{m}(\omega )\leq \lambda _{2}^{m}(\omega )\leq
\cdots \leq \lambda _{M_{m}}^{m}(\omega ) \)
be the increasingly ordered set of eigenvalues of \( W_{0mm}(\omega )
\),
\( m\in \N  \). Set\[
\Omegabad _{s}(k,n,m,i,j)=\{\omega \in \Omega _{s};\textrm{ }|\omega
k-\Delta _{mn}+\lambda ^{n}_{i}(\omega )-\lambda _{j}^{m}(\omega
)|<\psi (k,n,m)\}.\]
Then\[
\Omegabad _{s}=\bigcup _{(k,n,m)}\bigcup _{\substack {i,j\cr 1\leq
i\leq M_{n}\cr 1\leq j\leq M_{m}}}\Omegabad _{s}(k,n,m,i,j).\]

By assumption, if \( (k,n,m) \) is a non-critical index then \(
\Omegabad _{s}(k,n,m,i,j)=\emptyset  \)
(for any \( i,j \)). Further notice that, due to the symmetry
condition
(\ref{psi-symmetry-cond}), \( \Omegabad _{s}(k,n,m,i,j)=\Omegabad
_{s}(-k,m,n,j,i) \).

According to Lidskii Theorem (\cite{Kato}, Chap. II \S6.5), for any
\( j \), \( 1\leq j\leq M_{m} \), \( \lambda ^{m}_{j}(\omega
)-\lambda ^{m}_{j}(\omega ') \)
may be written as a convex combination (with non-negative
coefficients)
of eigenvalues of the operator \( W_{0mm}(\omega )-W_{0mm}(\omega ')
\).
Consequently,\[
\forall j,\textrm{ }1\leq j\leq M_{m},\textrm{ }\forall \omega
,\omega '\in \Omega _{s},\textrm{ }\omega \neq \omega ',\textrm{
}|\partial \lambda ^{m}_{j}(\omega ,\omega ')|\leq \| \partial
W_{0mm}(\omega ,\omega ')\| \leq \frac{1}{4}\, .\]

If \( \omega ,\omega '\in \Omegabad _{s}(k,n,m,i,j) \), \( \omega
\neq \omega ' \),
then \( (k,n,m) \) is necessarily a critical index
and\begin{eqnarray*}
\frac{2\psi (k,n,m)}{|\omega -\omega '|} & > & \left| \frac{(\omega
k-\Delta _{mn}+\lambda _{i}^{n}(\omega )-\lambda _{j}^{m}(\omega
))-(\omega 'k-\Delta _{mn}+\lambda _{i}^{n}(\omega ')-\lambda
_{j}^{m}(\omega '))}{\omega -\omega '}\right| \\
 & \geq  & |k|-\frac{1}{2}\, \geq \, \frac{1}{2}\, |k|\, .
\end{eqnarray*}
This implies that \( |\Omegabad _{s}(k,n,m,i,j)|\leq 4\psi
(k,n,m)/|k| \)
and so\begin{eqnarray*}
|\Omegabad _{s}| & \leq  & 2\sum _{\substack {(k,n,m)\cr k>0\cr
\frac{\Delta _{mn}}{2J}<k<\frac{2\Delta _{mn}}{J}}}\, \sum
_{\substack {i,j\cr 1\leq i\leq M_{n}\cr 1\leq j\leq
M_{m}}}\frac{4}{k}\, \psi (k,n,m)\, .
\end{eqnarray*}
This immediately leads to the desired inequality
(\ref{Omegabads_leq_sums_psi}).
\end{proof}

\section{\label{Sec:ConstructSeqs}Construction of the sequences
\protect\( \{\Omega _{s}\}\protect \)
and \protect\( \{A_{s}\}\protect \)}

For a non-diagonal multi-index \( (k,n,m) \) and \( s\in \Z _{+} \)
set \\
\begin{eqnarray}\label{psi_s-def}
\psi_s(k,n,m) & = &
\frac{1}{2}\,\Delta_{0}\phantom{IAAAAAAAAAAAAA}\quad
\textrm{if }(k,n,m)\textrm{ is non-critical and }k=0,\nonumber\\
& = &
\frac{7}{18}\,J\left(|k|-\frac{1}{2}\right)\quad\phantom{EAAAAAA}
\textrm{if }(k,n,m)\textrm{ is non-critical and }k\neq0,\nonumber\\
&=& \varphi_{s+1}
\left(\min\{M_{m},M_{n}\}\right)^{1/2}|k|^{1/2}e^{-\varrho_{s}|k|/2}\quad
\phantom{AAAA}\textrm{if }(k,n,m)\textrm{ is critical},\nonumber\\
&&
\end{eqnarray} where\[
\varrho_{s}=\frac{1}{E_{s}}-\frac{1}{E_{s+1}}.\]
Observe that \( \psi _{s} \) obeys the symmetry condition
(\ref{psi-symmetry-cond}).
The choice of \( \psi _{s}(k,n,m) \) for a non-critical index \(
(k,n,m) \)
was guided by the following lemma.

\lemcount

\begin{lem}
\label{Lem:noncrit_ind_notinterlaced}If \( \omega \in \Omega
_{s}\subset [\, \frac{8}{9}J,\frac{9}{8}J\, ] \),
\( (k,n,m)\in \Z \times \N \times \N  \) is a non-critical index
and \( W\in \gX _{s} \) satisfies\begin{equation}
\label{W_0mm_W_0nn_leq_RHS}
\| W_{0mm}(\omega )\| ,\| W_{0nn}(\omega )\| \leq \min
\left\{ \frac{1}{4}\, \Delta _{0},\frac{7}{72}\, J\right\}
\end{equation}
then the spectra \( \Spec (k\omega -\Delta _{mn}+W_{0nn}(\omega )) \),
\( \Spec (W_{0mm}(\omega )) \) are not interlaced (i.e., they are
separated by a real point \( p \) such that one of them lies below
and the other above \( p \)) and it holds\[
\dist \left( \Spec (k\omega -\Delta _{mn}+W_{0nn}(\omega )),\, \Spec
(W_{0mm}(\omega ))\right) \geq \psi (k,n,m).\]

\end{lem}
\begin{proof}
We distinguish two cases. If \( k\neq 0 \) then\[
|k\omega -\Delta _{mn}|=|k|\left| \omega -\frac{\Delta
_{mn}}{k}\right| \geq \frac{7}{18}\, J|k|\]
since, by assumption,\[
\frac{\Delta _{mn}}{k}-\omega \in \, ]-\infty ,\frac{1}{2}\,
J-\frac{8}{9}\, J\, ]\cup [\, 2J-\frac{9}{8}\, J,+\infty [\, .\]
So the distance may be estimated from below by\[
\frac{7}{18}\, J|k|-\| W_{0nn}(\omega )\| -\| W_{0mm}(\omega
)\| \geq \frac{7}{18}\, J\left( |k|-\frac{1}{2}\right) .\]
If \( k=0 \) then a lower bound to the distance is simply given by\[
\Delta _{0}-\| W_{0nn}(\omega )\| -\| W_{0mm}(\omega )\|
\geq \frac{1}{2}\, \Delta _{0}.\]

\end{proof}
Next we specify the way we shall construct the decreasing sequence
of sets \( \{\Omega _{s}\}_{s=0}^{\infty } \). Let \( \Omega _{0}=[\,
\frac{8}{9}J,\frac{9}{8}J\, ] \).
If \( W_{s}\in \gX _{s} \) has been already defined then we introduce
\( \Omega _{s+1}\subset \Omega _{s} \) as the set of \( (W_{s},\psi
_{s}) \)--non-resonant
frequencies. Recall that the real Banach space \( \gX _{s} \) is
determined by the choice of data \( \varphi _{s} \), \( E_{s} \)
and \( \Omega _{s} \), as explained in Section
\ref{Sec:ChoiceDirectSeqBanachSps}.

As a next step let us consider , for \( s\in \Z _{+} \), \( \omega
\in \Omega _{s+1} \)
and a non-diagonal index \( (k,n,m) \), a commutation
equation,\begin{equation}
\label{commeq_X_eq_Y}
(k\omega -\Delta _{mn}+(W_{s})_{0nn}(\omega ))X-X\,
(W_{s})_{0mm}(\omega )=Y,
\end{equation}
with an unknown \( X\in \BB (\HH _{m},\HH _{n}) \) and a right hand
side \( Y\in \BB (\HH _{m},\HH _{n}) \). Since \( \omega  \) is
\( (W_{s},\psi _{s}) \)--non-resonant the spectra \( \Spec (k\omega
-\Delta _{mn}+(W_{s})_{0nn}(\omega )) \)
and \( \Spec ((W_{s})_{0mm}(\omega )) \) don't intersect and so a
solution \( X \) exists and is unique. This way one can introduce
a linear mapping\[
(\Gamma _{s})_{knm}(\omega ):\BB (\HH _{m},\HH _{n})\to \BB (\HH
_{m},\HH _{n})\]
such that \( X=(\Gamma _{s})_{knm}(\omega )Y \) solves
(\ref{commeq_X_eq_Y}).
Moreover, according to Appendix A,\begin{equation}
\label{normGamma_leq-interlaced}
\| (\Gamma _{s})_{knm}(\omega )\| \leq \frac{\left( \min
\{M_{m},M_{n}\}\right) ^{1/2}}{\psi (k,n,m)}
\end{equation}
in the general case, and provided the spectra \( \Spec (k\omega
-\Delta _{mn}+(W_{s})_{0nn}(\omega )) \)
and

\( \Spec ((W_{s})_{0mm}(\omega )) \) are not interlaced it even
holds that\begin{equation}
\label{normGamma_leq-notinterlaced}
\| (\Gamma _{s})_{knm}(\omega )\| \leq \frac{1}{\psi (k,n,m)}\,
.
\end{equation}
From the uniqueness it is clear that \( \Ker ((\Gamma
_{s})_{knm}(\omega ))=0 \).

We extend the definition of \( (\Gamma _{s})_{knm} \) to diagonal
indices by letting \( (\Gamma _{s})_{0nn}(\omega )=0\in \BB (\BB (\HH
_{n},\HH _{n})) \).
This way we get an element\begin{equation}
\label{Gamma_s_elem_Map}
\Gamma _{s}\in \Map \left( \Omega _{s+1}\times \Z \times \N \times \N
,\sum _{n\in \N }\sideset {}{^{\oplus }}\sum _{m\in \N }\BB (\BB (\HH
_{m},\HH _{n}))\right) ,
\end{equation}
which naturally defines a linear mapping, denoted for simplicity by
the same symbol, \( \Gamma _{s}:{^{0}\gX }_{s}\to {^{0}\gX }_{s+1} \),
according to the rule\[
\Gamma _{s}(Y)_{knm}(\omega ):=(\Gamma _{s})_{knm}(\omega
)(Y_{knm}(\omega ))\, .\]
\lemcount

\begin{lem}
\label{Lem:estim_norm_Gamma_s}Assume that for all non-diagonal indices
\( (k,n,m) \) and \( \omega ,\omega '\in \Omega _{s+1} \), \( \omega
\neq \omega ' \),
it holds\begin{equation}
\label{part_Gamma_in_leq_k_plus_half}
\| \partial (\Gamma _{s})_{knm}^{-1}(\omega ,\omega ')\| \leq
|k|+\frac{1}{2},
\end{equation}
if \( \omega \in \Omega _{s+1} \) and \( (k,n,m) \) is a non-critical
index then the spectra \( \Spec (k\omega -\Delta
_{mn}+(W_{s})_{0nn}(\omega )) \)
and \( \Spec ((W_{s})_{0mm}(\omega )) \) are not interlaced
and\begin{equation}
\label{vphi_s_plus_1_leq_min()}
\varphi _{s+1}\leq \min \left\{ \frac{2}{3}\, \Delta
_{0},\frac{1}{6}\, J\right\} .
\end{equation}
Then the following upper estimate on the norm of \( \Gamma _{s}\in
\BB ({^{0}\gX }_{s},{^{0}\gX }_{s+1}) \)
holds true:\[
\| \Gamma _{s}\| \leq \frac{5}{2\, \varphi _{s+1}}\, .\]

\end{lem}
\begin{proof}
To estimate \( \| \Gamma _{s}\|  \) we shall use relation
(\ref{nGammaI})
of Proposition \ref{Prop:nGammaI} in Appendix B. Note
that\begin{eqnarray}
\| \partial (\Gamma _{s})_{knm}(\omega ,\omega ')\|  & = &
\| (\Gamma _{s})_{knm}(\omega )\, \partial (\Gamma
_{s})_{knm}^{-1}(\omega ,\omega ')\, (\Gamma _{s})_{knm}(\omega
')\| \nonumber \\
 & \leq  & \| (\Gamma _{s})_{knm}(\omega )\| \| (\Gamma
_{s})_{knm}(\omega ')\| \left( |k|+\frac{1}{2}\right) \,
.\label{partGamma}
\end{eqnarray}
If \( (k,n,m) \) is critical then we have, according to
(\ref{normGamma_leq-interlaced})
and (\ref{psi_s-def}),\[
\| (\Gamma _{s})_{knm}(\omega )\| \leq \frac{1}{\varphi
_{s+1}|k|^{1/2}}\, e^{\varrho_{s}|k|/2}\]
and consequently\begin{eqnarray*}
 &  & e^{-\varrho_{s}|k|}\left( \| (\Gamma _{s})_{knm}(\omega
)\| +\varphi _{s+1}\| \partial (\Gamma _{s})_{knm}(\omega
,\omega ')\| \right) \\
 &  & \qquad \qquad \leq \, e^{-\varrho_{s}|k|}\left(
\frac{1}{\varphi _{s+1}|k|^{1/2}}\,
e^{\varrho_{s}|k|/2}+\frac{|k|+\frac{1}{2}}{\varphi _{s+1}|k|}\,
e^{\varrho_{s}|k|}\right) \\
 &  & \qquad \qquad \leq \, \frac{1}{\varphi _{s+1}}\, \left(
1+1+\frac{1}{2|k|}\right) \, \leq \, \frac{5}{2\, \varphi _{s+1}}.
\end{eqnarray*}
If \( (k,n,m) \) is non-critical and \( k\neq 0 \) then we have,
according to (\ref{normGamma_leq-notinterlaced}) and
(\ref{psi_s-def}),\begin{eqnarray*}
\| (\Gamma _{s})_{knm}(\omega )\|  & \leq  & \frac{18}{7J\left(
|k|-\frac{1}{2}\right) }
\end{eqnarray*}
and consequently\begin{eqnarray*}
 &  & e^{-\varrho_{s}|k|}\left( \| (\Gamma _{s})_{knm}(\omega
)\| +\varphi _{s+1}\| \partial (\Gamma _{s})_{knm}(\omega
,\omega ')\| \right) \\
 &  & \qquad \qquad \leq \, \frac{18}{7J\left( |k|-\frac{1}{2}\right)
}\left( 1+\varphi _{s+1}\, \frac{18\left( |k|+\frac{1}{2}\right)
}{7J\left( |k|-\frac{1}{2}\right) }\right) \\
 &  & \qquad \qquad \leq \, \frac{1}{\varphi
_{s+1}}\frac{1}{6}\frac{36}{7}\left( 1+\frac{1}{6}\frac{54}{7}\right)
\, <\, \frac{2}{\varphi _{s+1}}\, .
\end{eqnarray*}
In the case when \( (k,n,m) \) is non-critical and \( k=0 \) one
gets similarly \( \| (\Gamma _{s})_{knm}(\omega )\| \leq
2/\Delta _{0} \)
and\begin{eqnarray*}
 &  & e^{-\varrho_{s}|k|}\left( \| (\Gamma _{s})_{knm}(\omega
)\| +\varphi _{s+1}\| \partial (\Gamma _{s})_{knm}(\omega
,\omega ')\| \right) \\
 &  & \qquad \qquad \leq \, \frac{2}{\Delta _{0}}\left( 1+\varphi
_{s+1}\, \frac{1}{\Delta _{0}}\right) \, \leq \, \frac{1}{\varphi
_{s+1}}\frac{4}{3}\left( 1+\frac{2}{3}\right) \, <\, \frac{5}{2\,
\varphi _{s+1}}\, .
\end{eqnarray*}

\end{proof}
Now we are able to specify the mappings \( \Theta _{u}^{s} \).
Set\begin{equation}
\label{As-def}
A_{s}=\Gamma _{s}\big ((1-\DD _{s})(W_{s}-\iota _{s-1}(W_{s-1}))\big
)\in {^{0}\gX }_{s+1}.
\end{equation}
\( W_{s}\in \gX _{s} \) satisfies (\ref{X-Hermitian}) and thus one
finds, when taking Hermitian adjoint of (\ref{commeq_X_eq_Y}), that\[
\left( (\Gamma _{s})_{knm}(\omega )Y\right) ^{\ast }=-(\Gamma
_{s})_{-k,m,n}(\omega )(Y^{\ast })\, .\]
This implies that \( A_{s} \) obeys condition
(\ref{As_anti-Hermitian}).
The mappings \( \Theta _{u}^{s} \), \( s<u \), are defined by equality
(\ref{0Theta_eq_comm_As_X}) (see also the comment following the
equality).

\section{\label{Sec:ProofMainTheorem}Proof of Theorem
\ref{MainTheorem}}

We start from the specification of the sequences \( \{\varphi _{s}\}
\)
and \( \{E_{s}\} \),\begin{equation}
\label{vphi_s_E_s-def}
\varphi _{s}=a\, s^{\alpha }q^{-rs}\textrm{ for }s\geq 1,\textrm{
}E_{s}=q^{s+1}\textrm{ for }s\geq 0,
\end{equation}
where \( \alpha >1 \) and \( q>1 \) are constants that are arbitrary
except of the restrictions\begin{equation}
\label{q-sigma_zeta(alpha)_eq_RHS}
q^{r}\geq e^{\alpha }\quad \textrm{and}\quad q^{-r}\zeta (\alpha
)\leq 3\, \ln 2
\end{equation}
(\( \zeta  \) stands for the Riemann zeta function),
and\begin{equation}
\label{a-def}
a=45\, e\, q^{2r}\epsilon _{V}.
\end{equation}
For example, \( \alpha =2 \) and \( q^{r}=e^{2} \) will do. The
value of \( \varphi _{0}\geq \varphi _{1}=a\, q^{-r} \) doesn't
influence
the estimates which follow, and we automatically have \( E_{-1}=1 \)
(this is a convenient convention). Condition \( r\, \ln (q)\geq
\alpha  \)
guarantees that the sequence \( \{\varphi _{s}\} \) is decreasing.
Note also that\[
\varrho_{s}=\frac{1}{E_{s}}-\frac{1}{E_{s+1}}=\left(
1-\frac{1}{q}\right) q^{-s-1}.\]

Another reason for the choice (\ref{vphi_s_E_s-def}) and (\ref{a-def})
is that the constants \( A_{\star } \), \( B_{\star } \) and \(
C_{\star } \),
as defined in (\ref{A0b0C0-def}), obey assumption
(\ref{hyp_estim_A0B0C0})
of Proposition \ref{Prop: A0B0C0_imply_Cor3}. Particularly, a
constraint
on the choice of \( \{\varphi _{s}\} \) and \( \{E_{s}\} \), namely
\( \sum ^{\infty }_{s=0}1/(\varphi _{s+1}(E_{s-1})^{r})<\infty  \),
is imposed by requiring \( B_{\star } \) to be finite. However this
is straightforward to verify. Actually, the constants may now be
expressed
explicitly,\[
A_{\star }=\frac{5e\, q^{2r}}{a},\textrm{ }B_{\star }=\frac{5e\,
q^{r}}{a}\, \zeta (\alpha ),\textrm{ }C_{\star }=\frac{5e\,
q^{r}}{a}\, ,\]
and thus conditions (\ref{hyp_estim_A0B0C0}) mean that\begin{equation}
\label{eps_V-conditions}
\epsilon _{V}\frac{5e\, q^{r}}{a}\, \zeta (\alpha )\leq \frac{1}{3}\,
\ln 2,\textrm{ }\epsilon _{V}\frac{5e\, q^{2r}}{a}\, \phi \left(
\epsilon _{V}\frac{15e\, q^{r}}{a}\right) \leq \frac{1}{9}\, .
\end{equation}
The latter condition in (\ref{eps_V-conditions}) is satisfied since
the LHS is bounded from above by (c.f. (\ref{phi-def}))\[
\frac{1}{9}\, \phi \left( \frac{1}{3}\, q^{-r}\right) \leq
\frac{1}{9}\, \phi \left( \frac{1}{3}\right) =1-\frac{2}{3}\,
e^{1/3}<\frac{1}{9}\, .\]
Concerning the former condition, the LHS equals \( q^{-r}\zeta
(\alpha )/9 \)
and so it suffices to chose \( \alpha  \) and \( q \) so that
(\ref{q-sigma_zeta(alpha)_eq_RHS})
is fulfilled. An additional reason for the choice
(\ref{vphi_s_E_s-def})
will be explained later.

Let us now summarise the construction of the sequences \( \{\gX
_{s}\} \),
\( \{W_{s}\} \) and \( \{\Theta _{u}^{s}\}_{s>u} \) which will finally
amount to a proof of Theorem \ref{MainTheorem}. Some more details
were already given in Section \ref{Sec:ConstructSeqs}. We set \(
\Omega _{0}=[\, \frac{8}{9}J,\frac{9}{8}J\, ] \)
and \( W_{0}=V_{0} \). Recall that the cut-offs \( V_{s} \) of \( V \)
were introduced in (\ref{cutoff_V}). In every step, numbered by \(
s\in \Z _{+} \),
we assume that \( \Omega _{t} \) and \( W_{t} \), with \( 0\leq t\leq
s \),
and \( A_{t} \), with \( 0\leq t\leq s-1 \), have already been
defined.
The mappings \( \Theta _{u}^{t} \), with \( u>t \), are given by
\( \Theta _{u}^{t}(X)=[\, \iota _{u,t+1}(A_{t}),X\, ] \) provided
\( A_{t}\in {^{0}\gX _{t+1}} \) satisfies condition
(\ref{As_anti-Hermitian}).
We define \( \Omega _{s+1}\subset \Omega _{s} \) as the set of \(
(W_{s},\psi _{s}) \)--non-resonant
frequencies, with \( \psi _{s} \) introduced in (\ref{psi_s-def}).
Consequently, the real Banach space \( \gX _{s+1} \) is defined as
well as its definition depends on the data \( \Omega _{s+1} \), \(
\varphi _{s+1} \)
and \( E_{s+1} \). Then we are able to introduce an element \( \Gamma
_{s} \)
(in the sense of (\ref{Gamma_s_elem_Map})) whose definition is based
on equation (\ref{commeq_X_eq_Y}) and which in turn determines a
bounded operator \( \Gamma _{s}\in \BB ({^{0}\gX _{s}},{^{0}\gX
_{s+1}}) \)
(with some abuse of notation). The element \( A_{s}\in {^{0}\gX
_{s+1}} \)
is given by equality (\ref{As-def}) and actually satisfies condition
(\ref{As_anti-Hermitian}). Knowing \( W_{t} \), \( t\leq s \),
and \( \Theta _{s+1}^{t} \), \( t\leq s \), (which is equivalent
to knowing \( A_{t} \), \( t\leq s \)) one is able to evaluate the
RHS of (\ref{recruleWs}) defining the element \( W_{s+1} \). Hence
one proceeds one step further.

We choose \( \epsilon _{\star }(r,\Delta _{0},J) \) maximal possible
so that\begin{equation}
\label{gammastar_bound1}
\frac{3e}{1-q^{-r}}\, \epsilon _{\star }(r,\Delta _{0},J)\leq \min
\left\{ \frac{1}{4}\, \Delta _{0},\frac{7}{72}\, J\right\}
\end{equation}
and\begin{equation}
\label{gammastar_bound2}
45\, e\, q^{r}\epsilon _{\star }(r,\Delta _{0},J)\leq \min \left\{
\frac{2}{3}\, \Delta _{0},\frac{1}{6}\, J\right\} \, .
\end{equation}
We claim that this choice guarantees that the construction goes
through.
Basically this means that \( \epsilon _{V}<\epsilon _{\star
}(r,\Delta _{0},J) \)
is sufficiently small so that all the assumptions occurring in the
preceding auxiliary results are satisfied in every step, with \( s\in
\Z _{+} \).
This concerns assumption (\ref{W_0mm_W_0nn_leq_RHS}) of Lemma
\ref{Lem:noncrit_ind_notinterlaced},\begin{equation}
\label{MTh_tobeverified_assumA}
\| (W_{s})_{0mm}(\omega )\| \leq \min \left\{ \frac{1}{4}\,
\Delta _{0},\frac{7}{72}\, J\right\} ,\quad \forall \omega \in \Omega
_{s},\textrm{ }\forall m\in \N ,
\end{equation}
assumption (\ref{part_W_0mm_leq_1_over_4}) of Lemma
\ref{Lem:estim_Omegabad_s},\begin{equation}
\label{MTh_tobeverified_assumB}
\| \partial (W_{s})_{0mm}(\omega ,\omega ')\| \leq
\frac{1}{4},\quad \forall \omega ,\omega '\in \Omega _{s},\textrm{
}\omega \neq \omega ',\textrm{ }\forall m\in \N ,
\end{equation}
assumptions (\ref{part_Gamma_in_leq_k_plus_half}) and
(\ref{vphi_s_plus_1_leq_min()})
of Lemma \ref{Lem:estim_norm_Gamma_s},\begin{equation}
\label{MTh_tobeverified_assumC}
\| \partial (\Gamma _{s})_{knm}^{-1}(\omega ,\omega ')\| \leq
|k|+\frac{1}{2},\quad \forall (k,n,m),\textrm{ }\forall \omega
,\omega '\in \Omega _{s},\textrm{ }\omega \neq \omega ',
\end{equation}
 and\begin{equation}
\label{MTh_tobeverified_assumD}
\varphi _{s+1}\leq \min \left\{ \frac{2}{3}\, \Delta
_{0},\frac{1}{6}\, J\right\} ,
\end{equation}
and assumption (\ref{normAs_leq_RHS}) of Proposition
\ref{Prop:diag_of_K_p_V},\begin{equation}
\label{MTh_tobeverified_assumE}
\| A_{s-1}\| \leq \frac{5}{2\varphi _{s}}\, \| W_{s-1}-\iota
_{s-2}(W_{s-2})\| \, .
\end{equation}

We can immediately do some simplifications. As the sequence \(
\{\varphi _{s}\} \)
is non-increasing condition (\ref{MTh_tobeverified_assumD}) reduces
to the case \( s=0 \). Since \( \varphi _{1}=45\, e\, q^{r}\epsilon
_{V} \)
the upper bound (\ref{gammastar_bound2}) implies
(\ref{MTh_tobeverified_assumD}).

Note also that (\ref{MTh_tobeverified_assumC}) is a direct consequence
of (\ref{MTh_tobeverified_assumB}). Actually, one deduces from the
definition of \( (\Gamma _{s})_{knm}(\omega ) \) (based on equation
(\ref{commeq_X_eq_Y})) that, \( \forall Y\in \BB (\HH _{m},\HH _{n})
\),\[
(\Gamma _{s})^{-1}_{knm}(\omega )Y=(k\omega -\Delta
_{mn}+(W_{s})_{0nn}(\omega ))Y-Y(W_{s})_{0mm}(\omega )\, .\]
Hence\[
\partial (\Gamma _{s})^{-1}_{knm}(\omega ,\omega ')Y=(k+\partial
(W_{s})_{0nn}(\omega ,\omega '))\, Y-Y\, \partial
(W_{s})_{0mm}(\omega ,\omega ')\]
and, assuming (\ref{MTh_tobeverified_assumB}),\[
\| \partial (\Gamma _{s})^{-1}_{knm}(\omega ,\omega ')\| \leq
|k|+\| \partial (W_{s})_{0nn}(\omega ,\omega '))\| +\|
\partial (W_{s})_{0mm}(\omega ,\omega ')\| \leq |k|+\frac{1}{2}\,
.\]

Let us show that in every step, with \( s\in \Z _{+} \), conditions
(\ref{MTh_tobeverified_assumA}), (\ref{MTh_tobeverified_assumB})
and (\ref{MTh_tobeverified_assumE}) are actually fulfilled. For \(
s=0 \),
condition (\ref{MTh_tobeverified_assumE}) is empty and condition
(\ref{MTh_tobeverified_assumB}) is obvious since \( W_{0}=V_{0} \)
doesn't depend on \( \omega  \). Condition
(\ref{MTh_tobeverified_assumA})
is obvious as well due to assumption (\ref{gammastar_bound1}) and
the fact that \( \| (W_{0})_{0mm}(\omega )\| =\|
(V_{0})_{0mm}\| \leq \epsilon _{V} \).

Assume now that \( t\in \Z _{+} \) and conditions
(\ref{MTh_tobeverified_assumA}),
(\ref{MTh_tobeverified_assumB}) and (\ref{MTh_tobeverified_assumE})
are satisfied in each step \( s\leq t \). Recall that in
(\ref{F_v_defby_phi_eps_E})
we have set \( F_{s}=5/\varphi _{s+1} \) and \( v_{s}=e\, \epsilon
_{V}/(E_{s-1})^{r} \).
We also keep the notation \( w_{s}=\| W_{s}-\iota
_{s-1}(W_{s-1})\| _{s} \),
with the convention \( W_{-1}=0 \).

We start with condition (\ref{MTh_tobeverified_assumE}). Using the
induction hypothesis, Lemma \ref{Lem:noncrit_ind_notinterlaced} and
Lemma \ref{Lem:estim_norm_Gamma_s} one finds that \( \| \Gamma
_{t}\| \leq F_{t}/2 \)
and so \( \| A_{t}\| \leq \| \Gamma _{t}\| \|
W_{t}-\iota _{t-1}(W_{t-1})\| \leq F_{t}w_{t}/2 \)
(c.f. (\ref{As-def}) and (\ref{normDsleq1})).

By the induction hypothesis and the just preceding step, \( \|
A_{s}\| \leq F_{s}w_{s} \)
for all \( s\leq t \). As we already know the constants \( A_{\star }
\),
\( B_{\star } \) and \( C_{\star } \) fulfil (\ref{hyp_estim_A0B0C0})
and so the quantities \( A \), \( B \) and \( C \) given by \(
A=\epsilon _{V}A_{\star } \),
\( B=\epsilon _{V}B_{\star } \) and \( C=\epsilon _{V}C_{\star } \)
(c.f. (\ref{ABC_defby_A0B0C0})) obey (\ref{B_leq_Aphi()_leq}) and
consequently inequality (\ref{ABd_ineq}) with \( d=3 \). By the
very choice of \( A \), \( B \) and \( C \) (c.f. (\ref{A0b0C0-def})
and (\ref{F_v_defby_phi_eps_E})) the quantities also obey relations
(\ref{Thmconv_assumA}), (\ref{Thmconv_assumB}) and
(\ref{Thmconv_assumC}).
This means that all assumptions of Proposition \ref{Prop:w_leq_dv}
are fulfilled for \( s\leq t \) (recall that \( \| \Theta
_{u}^{s}\| \leq 2\| A_{s} \)).
One easily finds that the conclusion of Proposition
\ref{Prop:w_leq_dv},
namely \( w_{s}\leq d\, v_{s} \), holds as well for all \( s \),
\( s\leq t+1 \). Clearly, \( \| (W_{s})_{0mm}(\omega )\| \leq
\| W_{s}\| _{s} \)
for all \( s \), and\[
\| W_{t+1}\| _{t+1}\leq \sum _{s=0}^{t+1}w_{s}\leq 3\sum
_{s=0}^{\infty }v_{s}=3e\, \epsilon _{V}\sum _{s=0}^{\infty
}q^{-rs}=\frac{3e}{1-q^{-r}}\, \epsilon _{V}.\]
By (\ref{gammastar_bound1}) we conclude that
(\ref{MTh_tobeverified_assumA})
is true for \( s=t+1 \).

Finally, using once more that \( w_{s}\leq 3v_{s} \) for \( s\leq t+1
\),\begin{eqnarray*}
\| \partial (W_{t+1})_{0mm}(\omega ,\omega ')\|  & \leq  & \sum
_{s=0}^{t+1}\| \partial \big (W_{s}-\iota _{s-1}(W_{s-1})\big
)_{0mm}(\omega ,\omega ')\| \\
 & \leq  & \sum _{s=0}^{t+1}\frac{1}{\varphi _{s}}\, \|
W_{s}-\iota _{s-1}(W_{s-1})\| _{s}\\
 & \leq  & \sum _{s=0}^{\infty }\frac{3v_{s}}{\varphi _{s+1}}\, .
\end{eqnarray*}
However, the last sum equals (c.f. (\ref{F_v_defby_phi_eps_E}) and
(\ref{B_leq_Aphi()_leq}))\[
\frac{3}{5}\sum ^{\infty }_{s=0}F_{s}v_{s}=\frac{3}{5}\, B\leq
\frac{1}{5}\, \ln 2<\frac{1}{4}\, .\]
This verifies (\ref{MTh_tobeverified_assumB}) for \( s=t+1 \) and
hence the verification of conditions (\ref{MTh_tobeverified_assumA}),
(\ref{MTh_tobeverified_assumB}) and (\ref{MTh_tobeverified_assumE})
is complete.

Set, as before, \( \Omega _{\infty }=\bigcap _{s=0}^{\infty }\Omega
_{s} \).
Next we are going to estimate the Lebesgue measure of \( \Omega
_{\infty } \),\[
|\Omega _{\infty }|=|\Omega _{0}|-|\Omega _{0}\setminus \Omega
_{\infty }|=\frac{17}{72}\, J-\sum _{s=0}^{\infty }|\Omega
_{s}\setminus \Omega _{s+1}|=\frac{17}{72}\, J-\sum _{s=0}^{\infty
}|\Omegabad _{s}|\, .\]
Recalling Lemma \ref{Lem:estim_Omegabad_s} jointly with Lemma
\ref{Lem:noncrit_ind_notinterlaced}
showing that the assumptions of Lemma \ref{Lem:estim_Omegabad_s}
are satisfied, and the explicit form of \( \psi  \) (\ref{psi_s-def})
we obtain\begin{eqnarray*}
|\Omegabad _{s}| & \leq  & 8\varphi _{s+1}\sum _{\substack {m,n\in \N
,\cr \Delta _{mn}\, >\, \frac{1}{2}J}}\mu _{mn}\, \sum _{\substack
{k\in \N ,\cr \max \{1,\frac{\Delta _{mn}}{2J}\}<k<\frac{2\Delta
_{mn}}{J}}}k^{-1/2}e^{-\varrho_{s}k/2}\\
 & \leq  & 8\varphi _{s+1}\sum _{\substack {m,n\in \N ,\cr \Delta
_{mn}\, >\, \frac{1}{2}J}}\mu _{mn}\frac{2\Delta _{mn}}{J}\left(
\frac{\Delta _{mn}}{2J}\right) ^{-1/2}e^{-\varrho_{s}\Delta
_{mn}/4J}\\
 & = & 32\, (2J)^{\sigma }\varphi _{s+1}\sum _{\substack {m,n\in \N
,\cr \Delta _{mn}\, >\, \frac{1}{2}J}}\frac{\mu _{mn}}{(\Delta
_{mn})^{\sigma }}\left( \frac{\Delta _{mn}}{2J}\right) ^{\sigma
+\frac{1}{2}}e^{-\varrho_{s}\Delta _{mn}/4J}\\
 & \leq  & 32\, 2^{\sigma }\varphi _{s+1}\left( \frac{2\sigma
+1}{e\varrho_{s}}\right) ^{\sigma +\frac{1}{2}}\Delta _{\sigma }(J)
\end{eqnarray*}
where we have used that if \( \alpha >0 \) and \( \beta >0 \) then
\( \sup _{x>0}x^{\alpha }e^{-\beta x}=(\frac{\alpha }{e\beta
})^{\alpha } \).
To complete the estimate we need that the sum \( \sum _{s=0}^{\infty
}\varphi _{s+1}/(\varrho_{s})^{\sigma +\frac{1}{2}} \)
should be finite which imposes another restriction on the choice of
\( \{\varphi _{s}\} \) and \( \{E_{s}\} \). With our choice
(\ref{vphi_s_E_s-def})
this is guaranteed by the condition \( r>\sigma +\frac{1}{2} \) since
in that case\[
\sum _{s=0}^{\infty }\frac{\varphi _{s+1}}{(\varrho_{s})^{\sigma
+\frac{1}{2}}}=\frac{a}{\left( 1-\frac{1}{q}\right) ^{\sigma
+\frac{1}{2}}}\sum _{s=0}^{\infty }(s+1)^{\alpha }q^{-(r-\sigma
-\frac{1}{2})(s+1)}<\infty \, .\]
Hence\begin{equation}
\label{formula_delta_1}
|\Omega _{\infty }|\geq \frac{17}{72}\, J-\delta _{1}(\sigma ,r)\,
\Delta _{\sigma }(J)\, \epsilon _{V}
\end{equation}
where\begin{equation}
\label{formula_delta_2}
\delta _{1}(\sigma ,r)=1440\, e\, q^{2r}2^{\sigma }\left(
\frac{2\sigma +1}{\left( 1-\frac{1}{q}\right) e}\right) ^{\sigma
+\frac{1}{2}}\Li _{-\alpha }(q^{-r+\sigma +\frac{1}{2}})
\end{equation}
Here \( \Li _{n}(z)=\sum _{k=1}^{\infty }z^{k}/k^{n} \) (\( |z|<1 \))
is the polylogarithm function. This shows
(\ref{measureJinfty_geq_RHS}).

To finish the proof let us assume that \( \omega \in \Omega _{\infty
} \).
We wish to apply Proposition \ref{Prop:diag_of_K_p_V}. Going through
its assumptions one finds that it only remains to make a note
concerning
equality (\ref{komega-Delta_As_eq_RHS}). In fact, this equality is
a direct consequence of the construction of \( A_{s}\in {^{0}\gX
_{s+1}} \).
Actually, by the definition of \( A_{s} \) (c.f. (\ref{As-def})),
\( A_{s}=\Gamma _{s}\big ((1-\DD _{s})(W_{s}-\iota
_{s-1}(W_{s-1}))\big ) \),
which means that for any \( \omega \in \Omega _{s+1} \) and all
indices
\( (k,n,m) \), \begin{equation}
\begin{aligned}
\label{commeq_for_As_A}\big(k\omega-\Delta_{mn}+(W_{s})_{0nn}(\omega)\big)(A_{s})_{knm}(\omega)
-(A_{s})_{knm}(\omega)(W_{s})_{0mm}(\omega)\qquad\qquad\quad \\
=\,\big ((1-\DD _{s})(W_{s}-\iota _{s-1}(W_{s-1}))\big )_{knm}(\omega
).
\end{aligned}
\end{equation} On the other hand, by the definition of \( \Theta
_{u}^{s} \)
(c.f. (\ref{0Theta_eq_comm_As_X})) and the definition of \( \DD _{s}
\)
(c.f (\ref{D_s-def})), and since \( \omega \in \Omega _{\infty } \),
it holds true that, \( \forall u \), \( u>s \),\begin{equation}
\label{commeq_for_As_B}\begin{aligned}
\Theta_{u}^{s}(\iota_{us}\DD_{s}(W_{s}))_{knm}(\omega) &\, =\, \big(
[\,\iota _{u,s+1}(A_{s}),\iota _{us}\DD _{s}(W_{s})\, ]\big
)_{knm}(\omega) \\
& \,=\, (A_{s})_{knm}(\omega
)(W_{s})_{0mm}-(W_{s})_{0nn}(A_{s})_{knm}(\omega ).
\end{aligned}
\end{equation}A combination of (\ref{commeq_for_As_A}) and
(\ref{commeq_for_As_B})
gives (\ref{komega-Delta_As_eq_RHS}). We conclude that according
to Proposition \ref{Prop:diag_of_K_p_V} the operator \( \bK +\bV  \)
is unitarily equivalent to \( \bK +\bD (\bW ) \) and hence has a
pure point spectrum. This concludes the proof of Theorem
\ref{MainTheorem}.

\section{\label{Sec:ConcludingRemarks}Concluding remarks}

The backbone of the proof of Theorem \ref{MainTheorem} forms an
iterative
procedure loosely called here and elsewhere the quantum KAM method.
One of the improvements attempted in the present paper was a sort
of optimalisation of this method, particularly from the point of view
of assumptions imposed on the regularity of the perturbation \( V \).
In this final section we would like to briefly discuss this feature
by comparing our presentation to an earlier version of the method.
We shall refer to paper \cite{DuclosStovicek} but the main points
of the discussion apply as well to other papers including the original
articles \cite{Bellissard}, \cite{Combescure} where the quantum
KAM method was established. For the sake of illustration we use a
simple but basic model: \( H=\sum _{m\in \N }m^{1+\alpha }Q_{m} \),
i.e., \( h_{m}=m^{1+\alpha } \), with \( 0<\alpha \leq 1 \), and
\( \dim Q_{m}=1 \); thus \( \mu _{mn}=1 \) and any \( \sigma
>1/\alpha  \)
makes \( \Delta _{\sigma }(J) \) finite. The perturbation \( V \)
is assumed to fulfill (34) for a given \( r\geq 0 \).

According to Theorem \ref{MainTheorem}, \( r \) is required to satisfy
\( r>\sigma +1/2 \) which may be compared to reference \cite[Theorem
4.1]{DuclosStovicek}
where one requires\begin{equation}
\label{ConRems:r_1}
r>\br _{1}=4\sigma +6+\left[ \frac{(4\sigma +6)\sigma }{1+\sigma
}\right] +1.
\end{equation}
The reason is that the procedure is done in two steps in the earlier
version; in the first step preceding the iterative procedure itself
the so-called adiabatic regularisation is applied on \( V \) in order
to achieve a regularity in time and {}``space{}'' (by the spatial
part one means the factor \( \HH  \) in \( \KK =L^{2}([\, 0,T\,
],dt)\otimes \HH  \))
of the type\begin{equation}
\label{ConRems:r_2}
\exists r_{1},r_{2}>\br _{2}=4\sigma +6,\quad \sup
_{knm}|k|^{r_{1}}|n-m|^{r_{2}}|V_{knm}|<\infty .
\end{equation}
The adiabatic regularisation brings in the summand \( \left[
\frac{(4\sigma +6)\sigma }{1+\sigma }\right] +1 \).
In the present version both the adiabatic regularisation and condition
(\ref{ConRems:r_2}) are avoided. This is related to the choice of
the norm in the auxiliary Banach spaces \( \gX _{s} \),\[
\| X\| _{s}=\sup _{\omega \neq \omega '}\sup _{n}\sum
_{k,m}F_{s}(k,n,m)\left( |X_{knm}(\omega )|+\varphi _{s}|\partial
X_{knm}(\omega ,\omega ')|\right) .\]
In the earlier version the weights were chosen as \(
F_{s}(k,n,m):=\exp ((|k|+|n-m|)/E_{s}) \)
in order to compensate small divisors occurring in each step of the
iterative method. A more careful control of the small divisors in
the present version allows less restrictive weights, namely \(
F_{s}(k,n,m)=\exp (|k|/E_{s}) \).
In more detail, indices labelling the small divisors are located in
a critical subset of the lattice \( \Z \times \N \times \N  \).
Definition
(\ref{def:crit_indcs}) of the critical indices implies a simple
estimate,\[
|k|\leq |k|+|n-m|\leq |k|+|\Delta _{mn}|\leq |k|(1+2J),\]
which explains why we effectively have, in the present version, \(
r_{2}=0 \).

The second remark concerns Diophantine-like estimates of the small
divisors governed by the sequence \( \{\psi _{s}\} \). A bit
complicated
definition (\ref{psi_s-def}) is caused by the classification of the
indices into critical and non-critical ones. However only the critical
indices are of importance in this context and thus we can simplify,
for the purpose of this discussion, the definition of \( \psi _{s} \)
to\[
\psi _{s}=\gamma _{s}|k|^{1/2}e^{-\varrho_{s}|k|/2},\quad \varphi
_{s+1}\geq \gamma _{s}>0.\]
Let us compare it to the choice made in \cite{DuclosStovicek}, namely
\( \psi _{s}=\gamma _{s}|k|^{-\sigma } \). The factors \( \gamma _{s}
\)
then occur in some key estimates; let us summarise them. The norm
of the operators \( \Gamma _{s}:\gX _{s}\to \gX _{s+1} \) are
estimated
as\[
\| \Gamma _{s}\| \leq \const \, \frac{\varphi _{s+1}}{\gamma
_{s}^{\, 2}}\]
(this is shown in Lemma \ref{Lem:estim_norm_Gamma_s} but note that
in this lemma we have set \( \gamma _{s}=\varphi _{s+1} \)). Another
important condition is the convergence of the series\[
B_{\star }=\const \sum _{s=0}^{\infty }\frac{\varphi _{s+1}}{\gamma
_{s}^{\, 2}(E_{s-1})^{r}}<\infty \]
(c.f. (\ref{A0b0C0-def}) but there again \( \gamma _{s}=\varphi
_{s+1} \)).
Finally, the measure of the set of resonant frequencies, \( |\cup
_{s}\Omegabad _{s}| \),
is estimated by\[
\sum _{s=0}^{\infty }|\Omegabad _{s}|\leq \const \sum _{s=0}^{\infty
}\frac{\gamma _{s}}{\varrho_{s}^{\, \sigma +\frac{1}{2}}}<\infty
,\quad \varrho_{s}=\frac{1}{E_{s}}-\frac{1}{E_{s+1}}\]
(shown in the part of the proof of Theorem \ref{MainTheorem} preceding
relation (\ref{formula_delta_1})). We recall that \( E_{s} \) denotes
the width of the truncation of the perturbation \( V \) at step \( s
\)
of the algorithm (c.f. (\ref{cutoff_V})). These conditions restrict
the choice of the sequences \( \{E_{s}\} \) and \( \{\gamma _{s}\} \)
which may also be regarded as parameters of the procedure.
Specification
(\ref{vphi_s_E_s-def}) of these parameters, with \( \gamma
_{s}=\varphi _{s+1} \),
can be compared to a polynomial behaviour of \( E_{s} \) and \(
\gamma _{s} \)
in the variable \( s \) in \cite{DuclosStovicek} where one sets
\( \varphi _{s+1}\equiv 1 \) and\[
E_{s}=\const \, (s+1)^{\nu -1},\quad \nu >2,\qquad \gamma _{s}=\const
\, (s+1)^{-\mu },\quad \mu >1.\]
The latter definition finally leads to the bound on the order of
regularity
of \( V \)\[
r>\frac{(2\sigma +1)\nu +3}{\nu -1}\, .\]
Thus in that case the bound varies from \( r>4\sigma +5 \) (for \(
\nu \to 2+ \);
this contributes to \( \br _{1} \) in (\ref{ConRems:r_1})) to \(
r>2\sigma +1 \)
(\( \nu \to +\infty  \)). This shows why we have chosen here to
truncate
with exponential \( E_{s} \), see (\ref{vphi_s_E_s-def}).

In the last remark let us mention a consequence of the equality \(
\gamma _{s}=\varphi _{s+1} \).
The conditions for convergence of \( B_{\star } \) and \( \cup
_{s}\Omega _{s}^{\textrm{bad}} \)
become (notice that \( \varrho_{s}=\const /E_{s} \))\[
\sum _{s}\frac{1}{\varphi _{s+1}(E_{s-1})^{r}}<\infty \quad
\textrm{and}\quad \sum _{s}\varphi _{s+1}E_{s}^{\, \sigma
+\frac{1}{2}}<\infty \]
and are fulfilled for \( r>\sigma +\frac{1}{2} \). There is however
a drawback with this choice. Notice the role the coefficients \(
\varphi _{s} \)
play in the definition (\ref{norm_Xs-def}) of the norm \( \| \cdot
\| _{s} \).
Since \( \varphi _{s}\rightarrow 0 \) as \( s\rightarrow \infty  \)
one looses the control of the Lipschitz regularity in \( \omega  \)
in the limit of the iterative procedure. This means that we have no
information about the regularity of the eigenvectors and the
eigenvalues
of \( \bK +\bV  \) with respect to \( \omega  \). With \( r>2\sigma
+1 \)
we could have taken \( \varphi _{s+1}=1 \) and obtained that these
eigenvalues and vectors are indeed Lipschitz in \( \omega  \).

\section*{Appendix A. Commutation equation}

Suppose that \( \gX  \) and \( \gY  \) are Hilbert spaces, \( \dim
\gX <\infty  \),
\( \dim \gY <\infty  \), \( A\in \BB (\gY ) \), \( B\in \BB (\gX ) \),
both \( A \) and \( B \) are self-adjoint, and \( V\in \BB (\gX ,\gY
) \).
If \( \gamma  \) is a simple closed and positively oriented curve
in the complex plane such that \( \Spec (A) \) lies in the domain
encircled by \( \gamma  \) while \( \Spec (B) \) lies in its
complement
then the equation\begin{equation}
\label{AW-WB_eq_V}
AW-WB=V
\end{equation}
has a unique solution \( W\in \BB (\gX ,\gY ) \) given
by\begin{equation}
\label{inverse_comm_as_int}
W=\frac{1}{2\pi \i }\oint _{\gamma }(A-z)^{-1}V(B-z)^{-1}dz\, .
\end{equation}
The verification is straightforward.

Denote \( M_{1}=\dim \gX  \), \( M_{2}=\dim \gY  \). We shall need
the following two estimates on the norm of \( X\in \BB (\gX ,\gY )
\):\begin{eqnarray}
\| X\| ^{2} & \leq  & \sum ^{M_{2}}_{i=1}\sum
^{M_{1}}_{j=1}|X_{ij}|^{2}=\Tr X^{\ast }X\quad
(\textrm{Hilbert}-\textrm{Schmidt norm}),\label{estm_Xnorm_ab} \\
\| X\| ^{2} & \geq  & \max \left\{ \max _{1\leq i\leq
M_{2}}\sum ^{M_{1}}_{j=1}|X_{ij}|^{2},\max _{1\leq j\leq M_{1}}\sum
^{M_{2}}_{i=1}|X_{ij}|^{2}\right\} ,\label{estm_Xnorm_be}
\end{eqnarray}
where \( (X_{ij}) \) is a matrix of \( X \) expressed with respect
to any orthonormal bases in \( \gX  \) and \( \gY  \).

If \( \sup \Spec (A)<\inf \Spec (B) \) or \( \sup \Spec (B)<\inf
\Spec (A) \)
we shall say that \( \Spec (A) \) and \( \Spec (B) \) are not
interlaced.\propcount

\begin{prop}
If \( \Spec (A) \) and \( \Spec (B) \) are not interlaced then\[
\| W\| \leq \frac{\| V\| }{\dist (\Spec (A),\Spec (B))}\,
,\]
otherwise, if \( \Spec (A) \) and \( \Spec (B) \) don't intersect
but are interlaced,\[
\| W\| \leq (\min \left\{ \dim \gX ,\dim \gY \right\}
)^{1/2}\frac{\| V\| }{\dist (\Spec (A),\Spec (B))}\, .\]

\end{prop}
\begin{proof}
(1) If \( d=\inf \Spec (B)-\sup \Spec (A)>0 \) then, after a usual
limit procedure, we can choose for the integration path in
(\ref{inverse_comm_as_int})
the line which is parallel to the imaginary axis and intersects the
real axis in the point \( x_{0}=(\sup \Spec (A)+\inf \Spec (B))/2 \).
So\begin{eqnarray*}
\| W\| ^{2} & \leq  & \frac{1}{2\pi }\int ^{\infty }_{-\infty
}\| (A-x_{0}-\i s)^{-1}\| \| V\| \| (B-x_{0}-\i
s)^{-1}\| \, ds\\
 & = & \frac{\| V\| }{2\pi }\int ^{\infty }_{-\infty
}\frac{ds}{\left( \frac{d}{2}\right) ^{2}+s^{2}}\\
 & = & \frac{\| V\| }{d}\, .
\end{eqnarray*}

(2) In the interlaced case we choose orthonormal bases in \( \gX  \)
and \( \gY  \) so that \( A \) and \( B \) are diagonal, \( A=\diag
(a_{1},\dots ,a_{M_{2}}) \)
and \( B=(b_{1},\dots ,b_{M_{1}}) \). For brevity let us denote \(
\dist (\Spec (A),\Spec (B)) \)
by \( d \). Then \( W_{ij}=V_{ij}/(a_{i}-b_{j}) \), and we can use
(\ref{estm_Xnorm_ab}), (\ref{estm_Xnorm_be}) to
estimate\begin{eqnarray*}
\| W\| ^{2} & \leq  & \sum ^{M_{2}}_{i=1}\sum
^{M_{1}}_{j=1}\left| \frac{V_{ij}}{a_{i}-b_{j}}\right| ^{2}\leq \sum
^{M_{2}}_{i=1}\sum ^{M_{1}}_{j=1}\frac{|V_{ij}|^{2}}{d^{2}}\\
 & \leq  & \sum ^{M_{2}}_{i=1}\frac{\| V\|
^{2}}{d^{2}}=M_{2}\frac{\| V\| ^{2}}{d^{2}}\, .
\end{eqnarray*}
Symmetrically, \( \| W\| \leq M_{1}^{1/2}\| V\| /d \),
and the result
follows.
\end{proof}

\section*{Appendix B. Choice of a norm in a Banach space}

Let\[
\HH =\sideset {}{^{\oplus }}\sum _{n\in \N }\HH _{n}\]
be a decomposition of a Hilbert space into a direct sum of mutually
orthogonal subspaces, and \( \Omega \subset \R  \). To any couple
of positive real numbers, \( \varphi  \) and \( E \), we relate
a subspace\[
\gA \subset L^{\infty }\left( \Omega \times \Z \times \N \times \N
,\sum _{n\in \N }\sideset {}{^{\oplus }}\sum _{m\in \N }\BB (\HH
_{m},\HH _{n})\right) \]
formed by those elements \( \VV  \) which satisfy\[
\VV _{knm}(\omega )\in \BB (\HH _{m},\HH _{n})\]
and have finite norm\begin{equation}
\label{norm_AppB-def}
\| \VV \| =\sup _{\substack {\omega ,\omega '\in \Omega \cr
\omega \neq \omega '}}\, \sup _{n\in \N }\, \sum _{k\in \Z }\sum
_{m\in \N }\left( \| \VV _{knm}(\omega )\| +\varphi \, \|
\partial \VV _{knm}(\omega ,\omega ')\| \right) e^{|k|/E}
\end{equation}
where \( \partial  \) stands for the difference operator\[
\partial \VV (\omega ,\omega ')=\frac{\VV (\omega )-\VV (\omega
')}{\omega -\omega '}\, .\]
Note that the difference operator obeys the rule\begin{equation}
\label{partialLeibniz}
\partial (\UU \VV )(\omega ,\omega ')=\partial \UU (\omega ,\omega
')\, \VV (\omega ')+\UU (\omega )\partial \VV (\omega ,\omega ')\, .
\end{equation}
\propcount

\begin{prop}
The norm in \( \gA  \) is an algebra norm with respect to the
multiplication\begin{equation}
\label{prodUV}
(\UU \VV )_{knm}(\omega )=\sum _{\ell \in \Z }\sum _{p\in \N }\UU
_{k-\ell ,n,p}(\omega )\, \VV _{\ell pm}(\omega )\, .
\end{equation}

\end{prop}
\begin{proof}
We have to show that\begin{equation}
\label{algnorm}
\| \UU \VV \| \leq \| \UU \| \| \VV \| \, .
\end{equation}
For brevity let us denote (in this proof)\begin{eqnarray*}
\XX _{p}(\omega ) & = & \sum _{\ell \in \Z }\sum _{m\in \N }\| \VV
_{\ell pm}(\omega )\| \, e^{|\ell |/E},\\
\partial \XX _{p}(\omega ,\omega ') & = & \sum _{\ell \in \Z }\sum
_{m\in \N }\| \partial \VV _{\ell pm}(\omega ,\omega ')\| \,
e^{|\ell |/E}.
\end{eqnarray*}
Here \( \partial \XX  \) is an {}``inseparable{}'' symbol (which
this time doesn't have the meaning \( \partial  \) of \( \XX  \)).
It holds\begin{eqnarray*}
 &  & \sum _{k}\sum _{m}\| (\UU \VV )_{knm}(\omega )\| \,
e^{|k|/E}\\
 &  & \qquad \leq \sum _{k}\sum _{m}\sum _{\ell }\sum _{p}\| \UU
_{k-\ell ,n,p}(\omega )\| \, e^{|k-\ell |/E}\| \VV _{\ell
pm}(\omega )\| \, e^{|\ell |/E}\\
 &  & \qquad =\sum _{k}\sum _{m}\sum _{\ell }\sum _{p}\| \UU
_{knp}(\omega )\| \, e^{|k|/E}\| \VV _{\ell pm}(\omega )\|
\, e^{|\ell |/E}\\
 &  & \qquad =\sum _{k}\sum _{p}\| \UU _{knp}(\omega )\| \,
e^{|k|/E}\XX _{p}(\omega )\, .
\end{eqnarray*}
Similarly, using (\ref{partialLeibniz}),\begin{eqnarray*}
\sum _{k}\sum _{m}\| \partial (\UU \VV )_{knm}(\omega )\| \,
e^{|k|/E} & \leq  & \sum _{k}\sum _{m}\sum _{\ell }\sum _{p}\left(
\| \UU _{knp}(\omega )\| \, e^{|k|/E}\| \partial \VV _{\ell
pm}(\omega ,\omega ')\| \, e^{|\ell |/E}\right. \\
 &  & \\
 &  & \qquad \left. +\| \partial \UU _{knp}(\omega ,\omega ')\|
\, e^{|k|/E}\| \VV _{\ell pm}(\omega ')\| \, e^{|\ell
|/E}\right) \\
 & = & \sum _{k}\sum _{p}\left( \| \UU _{knp}(\omega )\| \,
\partial \XX _{p}(\omega ,\omega ')\right. \\
 &  & \qquad \left. +\| \partial \UU _{knp}(\omega ,\omega ')\|
\, \XX _{p}(\omega ')\right) e^{|k|/E}\, .
\end{eqnarray*}
A combination of these two inequalities gives\begin{eqnarray*}
 &  & \sum _{k}\sum _{m}\left( \| (\UU \VV )_{knm}(\omega )\|
+\varphi \, \| \partial (\UU \VV )_{knm}(\omega ,\omega ')\|
\right) e^{|k|/E}\\
 &  & \qquad \leq \sum _{k}\sum _{p}\left( \| \UU _{knp}(\omega
)\| (\XX _{p}(\omega )+\varphi \, \partial \XX _{p}(\omega ,\omega
'))+\varphi \, \| \partial \UU _{knp}(\omega ,\omega ')\| \,
\XX _{p}(\omega ')\right) e^{|k|/E}\\
 &  & \qquad \leq \sup _{\omega ,\omega '}\sup _{p}(\XX _{p}(\omega
)+\varphi \, \partial \XX _{p}(\omega ,\omega '))\sum _{k}\sum
_{p}(\| \UU _{knp}(\omega )\| +\varphi \, \| \partial \UU
_{knp}(\omega ,\omega ')\| )e^{|k|/E}\\
 &  & \qquad =\| \VV \| \, \sum _{k}\sum _{p}(\| \UU
_{knp}(\omega )\| +\varphi \, \| \partial \UU _{knp}(\omega
,\omega ')\| )e^{|k|/E}\, .
\end{eqnarray*}
To obtain (\ref{algnorm}) it suffices to apply \( \sup _{\omega
,\omega '}\sup _{n} \)
to this inequality.
\end{proof}
Suppose now that two couples of positive real numbers, \( (\varphi
_{1},E_{1}) \)
and \( (\varphi _{2},E_{2}) \), are given and that it
holds\begin{equation}
\label{condsonFw12}
\varrho=\frac{1}{E_{1}}-\frac{1}{E_{2}}\geq 0\quad \textrm{and}\quad
\varphi _{2}\leq \varphi _{1}\, .
\end{equation}
Consequently, we have two Banach spaces, \( \gA _{1} \) and \( \gA
_{2} \).
Furthermore, we suppose that there is given an element\begin{equation}
\label{Gammaintro}
\Gamma \in \Map \left( \Omega \times \Z \times \N \times \N ,\sum
_{n\in \N }\sideset {}{^{\oplus }}\sum _{m\in \N }\BB (\BB (\HH
_{m},\HH _{n}))\right) \, ,
\end{equation}
such that for each couple \( (\omega ,k)\in \Omega \times \Z  \)
and each double index \( (n,m)\in \N \times \N  \), \( \Gamma
_{knm}(\omega ) \)
belongs to \( \BB (\BB (\HH _{m},\HH _{n})) \). \( \Gamma  \)
naturally
determines a linear mapping, called for the sake of simplicity also
\( \Gamma  \), from \( \gA _{1} \) to \( \gA _{2} \), according
to the prescription\begin{equation}
\label{Gammamap}
\Gamma (\VV )_{knm}(\omega )=\Gamma _{knm}(\omega )(\VV _{knm}(\omega
))\, .
\end{equation}
Concerning the difference operator, in this case one can apply the
rule\[
\partial \left( \Gamma (\VV )\right) (\omega ,\omega ')=\partial
\Gamma (\omega ,\omega ')\, (\VV (\omega '))+\Gamma (\omega
)(\partial \VV (\omega ,\omega '))\, .\]
\propcount

\begin{prop}
\label{Prop:nGammaI}The norm of \( \Gamma :\gA _{1}\to \gA _{2} \)
can be estimated as follows,\begin{equation}
\label{nGammaI}
\| \Gamma \| \leq \sup _{\substack {\omega ,\omega '\in \Omega
\cr \omega \neq \omega '}}\, \sup _{k\in \Z }\, \sup _{(n,m)\in \N
\times \N }e^{-\varrho |k|}\left( \| \Gamma _{knm}(\omega )\|
+\varphi _{2}\, \| \partial \Gamma _{knm}(\omega ,\omega ')\|
\right) \, .
\end{equation}

\end{prop}
\begin{proof}
Notice that, if convenient, one can interchange \( \omega  \) and
\( \omega ' \) in \( \| \partial \UU (\omega ,\omega ')\|  \).
It
holds\begin{eqnarray*}
 &  & \sum _{k}\sum _{m}\left( \| \Gamma _{knm}(\omega )(\VV
_{knm}(\omega ))\| +\varphi _{2}\, \| \partial \left( \Gamma
_{knm}(\VV _{knm})\right) (\omega ,\omega ')\| \right)
e^{|k|/E_{2}}\\
 &  & \qquad \leq \sum _{k}\sum _{m}\left( \| \VV _{knm}(\omega
)\| (\| \Gamma _{knm}(\omega )\| +\varphi _{2}\, \|
\partial \Gamma _{knm}(\omega ,\omega ')\| )e^{-\varrho
|k|}\right. \\
 &  & \qquad \qquad \left. +\varphi _{2}\, \| \partial \VV
_{knm}(\omega ,\omega ')\| \, \| \Gamma _{knm}(\omega ')\|
e^{-\varrho |k|}\right) e^{|k|/E_{1}}\\
 &  & \qquad \leq \sup _{\omega ,\omega '}\, \sup _{k}\, \sup
_{(n,m)}e^{-\varrho |k|}\left( \| \Gamma _{knm}(\omega )\|
+\varphi _{2}\, \| \partial \Gamma _{knm}(\omega ,\omega ')\|
\right) \\
 &  & \qquad \qquad \times \sum _{k}\sum _{m}\left( \| \VV
_{knm}(\omega )\| +\varphi _{1}\, \| \partial \VV _{knm}(\omega
,\omega ')\| \right) e^{|k|/E_{1}}\, .
\end{eqnarray*}
To finish the proof it suffices to apply \( \sup _{\omega ,\omega
'}\, \sup _{n} \)
to this inequality.
\end{proof}

\end{document}